\documentclass[11pt,a4paper]{article}

\usepackage{a4}
\usepackage{graphics}
\usepackage{amsthm}
\usepackage{amstext}
\usepackage{amsmath}
\usepackage{amsbsy}
\usepackage{amssymb}
\usepackage{amsfonts}
\usepackage[vflt]{floatflt}
\usepackage{epsfig}
\def\ra{\rightarrow}
\newcommand\nutau{{\nu_\tau}}

\newcommand\numu{{\nu_\mu}}

\newcommand\nue{{\nu_e}}

\def\dm2{\Delta m^2}
\def\sq2{sin^2(2\Theta)}

\def\etal{{\it et\ al.}}

\def\NOE{{\em N\raise.5ex\hbox{O \kern-0.47em}E\kern.4em}}
\def\ICARUS{ICARUS\ }

\begin{document}


\begin{flushright}
LNGS-P28/2001\\
ICARUS-TM/2001-03\\
March 1, 2001\\
\end{flushright}

\vskip 0.5cm

\vskip 1.0cm
\begin{center}
{\bf \Large
THE ICARUS EXPERIMENT, 
A Second-Generation Proton
Decay Experiment
and
Neutrino Observatory at
the Gran Sasso Laboratory}
\end{center}

\setcounter{footnote}{0}   
\newcommand{\Aname}[2]{\large #1$^{\ref{#2}}$}
\newcommand{\Anamenew}[3]{\large #1$^{\ref{#2},\ref{#3}}$}
\def\titlefoot#1{\stepcounter{footnote}\footnotetext{#1}}

\def\A{\kern+.6ex\lower.42ex\hbox{$\scriptstyle \iota$}\kern-1.20ex a}
\def\E{\kern+.5ex\lower.42ex\hbox{$\scriptstyle \iota$}\kern-1.10ex e}

\def\Alfredo{$^\dagger$}
\begin{center}

by the ICARUS Collaboration\\

\Aname{F.~Arneodo}{LNGS},
\Aname{B.~Babussinov}{Padova},
\Aname{A.~Badertscher}{ETH},
\Aname{B.~Bade\l{}ek}{iepwar},
\Aname{G.~Battistoni}{Milano},
\Aname{P.~Benetti}{Pavia},
\Aname{E.~Bernardini}{LNGS},
\Aname{M.~Bischofberger}{ETH},
\Aname{A.~Borio~di~Tigliole}{Polimilano},
\Aname{R.~Brunetti}{Pavia},
\Aname{A.~Bueno}{ETH},
\Aname{E.~Calligarich}{Pavia},
\Aname{M.~Campanelli}{ETH},
\Aname{C.~Carpanese}{ETH},
\Aname{D.~Cavalli}{Milano},
\Aname{F.~Cavanna}{Aquila},
\Aname{P.~Cennini}{CERN},
\Aname{S.~Centro}{Padova},
\Aname{A.~Cesana}{Polimilano},
\Aname{C.~Chen}{CHINA},
\Aname{Y.~Chen}{CHINA},
\Aname{D.~Cline}{UCLA},
\Aname{A.~D\A browska}{niewodnicza},
\Aname{R.~Dolfini}{Pavia},
\Aname{M.~Felcini}{ETH},
\Anamenew{A.~Ferrari}{CERN}{Alfredo},
\Aname{K.~He}{CHINA},
\Aname{J.~Holeczek}{Katowice},
\Aname{X.~Huang}{CHINA},
\Aname{A.~Gigli~Berzolari}{Pavia},
\Aname{I.~Gil-Botella}{ETH},
\Aname{D.~Grech}{wroclaw},
\Aname{B.~Jokisz}{Katowice},
\Aname{C.~Juszczak}{wroclaw},
\Aname{J.~Kisiel}{Katowice},
\Aname{D.~Kie\l{}czewska}{iepwar},
\Aname{T.~Koz\l{}owski}{inswar},
\Aname{M.~Laffranchi}{ETH},
\Aname{J.~\L{}agoda}{iepwar},
\Aname{Z.~Li}{CHINA},
\Aname{F.~Lu}{CHINA},
\Aname{J.~Ma}{CHINA},
\Aname{C.~Matthey}{UCLA},
\Aname{F.~Mauri}{Pavia},
\Aname{M.~Markiewicz}{pntkrakow},
\Aname{D.~Mazza}{Aquila},
\Aname{L.~Mazzone}{Pavia},
\Aname{G.~Meng}{Padova},
\Aname{C.~Montanari}{Pavia},
\Aname{M.~Moszy\'nski}{inswar}, 
\Aname{S.~Navas-Concha}{ETH},
\Aname{G.~Nurzia}{Aquila},
\Aname{S.~Otwinowski}{UCLA},
\Aname{O.~Palamara}{LNGS},
\Aname{D.~Pascoli}{Padova},
\Aname{J.~Pasternak}{wroclaw},
\Aname{L.~Periale}{Torino},
\Aname{S.~Petrera}{Aquila},
\Aname{G.~Piano~Mortari}{Aquila},
\Aname{A.~Piazzoli}{Pavia},
\Anamenew{P.~Picchi}{Torino}{Picchi},
\Anamenew{F.~Pietropaolo}{CERN}{Pietropaolo},
\Aname{A.~Rappoldi}{Pavia},
\Aname{G.L.~Raselli}{Pavia},
\Aname{J.~Rico}{ETH},
\Aname{E.~Rondio}{inswar},
\Aname{M.~Rossella}{Pavia},
\Aname{C.~Rossi}{Aquila}
\Aname{A.~Rubbia}{ETH},
\Anamenew{C.~Rubbia}{Pavia}{spokesman},
\Aname{P.~Sala}{Milano},
\Aname{T.~Rancati}{Milano},
\Aname{D.~Scannicchio}{Pavia},
\Aname{F.~Sergiampietri}{Pisa},
\Aname{J.~Sobczyk}{wroclaw},
\Aname{N.~Sinanis}{ETH}
\Aname{J.~Stepaniak}{inswar},
\Aname{M.~Stodulski}{niewodnicza},
\Aname{M.~Szeptycka}{inswar}, 
\Aname{M.~Szleper}{inswar},
\Aname{M.~Terrani}{Polimilano}
\Aname{S.~Ventura}{Padova}
\Aname{C.~Vignoli}{Pavia},
\Aname{H.~Wang}{UCLA},
\Aname{M.~W\'ojcik}{jagell},
\Aname{J.~Woo}{UCLA},
\Aname{G.~Xu}{CHINA},
\Aname{Z.~Xu}{Pavia},
\Aname{A.~Zalewska}{niewodnicza},
\Aname{J.~Zalipska}{inswar},
\Aname{C.~Zhang}{CHINA},
\Aname{Q.~Zhang}{CHINA},
\Aname{S.~Zhen}{CHINA},
\Aname{W.~Zipper}{Katowice}

\titlefoot{Laboratori Nazionali di Gran Sasso, INFN,  s.s. 17bis,
km 18+910, Assergi (AQ), Italy \label{LNGS}} 
\titlefoot{Dipartimento di Fisica e INFN, Universit\`a di Padova, 
via Marzolo 8, Padova, Italy\label{Padova}}
\titlefoot{Institute for Particle Physics, ETH H\"onggerberg, Z\"urich, Switzerland\label{ETH}}
\titlefoot{Institute of Experimental Physics, Warsaw University, Warszawa, Poland\label{iepwar}}
\titlefoot{Dipartimento di Fisica e 
INFN, Universit\`a di Milano, via Celoria 16, Milano, Italy\label{Milano}}
\titlefoot{Dipartimento di Fisica e INFN, Universit\`a di Pavia, via Bassi 6, 
Pavia, Italy\label{Pavia}}
\titlefoot{Politecnico di Milano (CESNF), Universit\`a di Milano, 
via Ponzio 34/3, Milano, Italy\label{Polimilano}}
\titlefoot{Dipartimento d Fisica e INFN, Universit\`a dell'Aquila, 
via Vetoio, L'Aquila, Italy\label{Aquila}}
\titlefoot{CERN, CH 1211 Geneva 23, Switzerland\label{CERN}}
\titlefoot{IHEP -- Academia Sinica, 19 Yuqnan Road, Beijing,
People's Republic of China\label{CHINA}}
\titlefoot{Department of Physics, UCLA, Los Angeles, CA 90024, USA
\label{UCLA}}
\titlefoot{H.Niewodnicza\'nski Institute of Nuclear Physics, Krak\'ow, Poland\label{niewodnicza}}
\titlefoot{Institute of Physics, University of Silesia, Katowice, Poland\label{Katowice}}
\titlefoot{Institute of Theoretical Physics, Wroc\l{}aw University, Wroc\l{}aw, Poland\label{wroclaw}}
\titlefoot{A.So\l{}tan Institute for Nuclear Studies, Warszawa, Poland\label{inswar}}
\titlefoot{Faculty of Physics and Nuclear Techniques, University of Mining and 
Metallurgy, Krak\'ow, Poland\label{pntkrakow}}
\titlefoot{University of Torino, Torino, Italy \label{Torino}}
\titlefoot{INFN Pisa, via Livornese 1291, San Piero a Grado (PI),
Italy\label{Pisa}}
\titlefoot{Institute of Physics, Jagellonian University, Krak\'ow, Poland\label{jagell}}
\renewcommand{\thefootnote}{\fnsymbol{footnote}}
\setcounter{footnote}{1}
\titlefoot{Also at Dipartimento di
Fisica e INFN, Universit\`a di Milano, via Celoria 16, Milano, Italy\label{Alfredo}}
\titlefoot{Also at Laboratori Nazionali di Frascati,
INFN,  Frascati, Italy and Istituto di Cosmogeofisica, CNR, Torino, Italy\label{Picchi}}
\titlefoot{Also at Dipartimento di Fisica e INFN, Universit\`a di Padova, via Marzolo 8, Padova, 
Italy\label{Pietropaolo}}
\titlefoot{Spokesman\label{spokesman}}
\end{center}

\setcounter{footnote}{0}   

\begin{abstract}
The final phase of the ICARUS physics program
requires a sensitive mass of liquid Argon of 5000 tons or more. This is still true today, 
even after the operation of large or the planning of even larger underground detectors.
The superior bubble-chamber-like features 
of the ICARUS detector will always provide additional and fundamental 
contributions to the 
field.

The most conservative way to reach a liquid Argon sensitive mass of
5000~tons is to start with a first prototype of a modest mass: the T600 detector.
This step-wise strategy allowed us to develop progressively the necessary know-how 
to build a large liquid Argon detector.

The T600 detector stands today as the first 
living proof that such large detector can be built and that liquid Argon
imaging technology can be 
implemented on such large scales.

After the successful completion of a series of technical tests to be
performed at the assembly hall in Pavia, the T600 detector will be ready to
be transported into the LNGS tunnel.
The operation of the T600 at the LNGS will allow us (1) to develop the local 
infrastructure needed to operate our large detector (2) to start the
handling
of the underground liquid argon technology
(3) to study the local background 
(4) to start the data taking with an initial liquid argon mass that
will reach in a 5-6 year program the multi-kton goal.
The T600 is to be considered as the first 
milestone on the road towards a total sensitive mass of 5000 tons: it is
the first piece of the detector to be complemented by further modules 
of appropriate size and dimensions,
in
order to reach in a most efficient and rapid way the final design mass.

In this document, we describe the physics program that will be
accomplished within the first phase of the program.
\end{abstract}

\section{Introduction}

The ICARUS physics program has been described in Volume I of the 1994 
proposal~\cite{Cennini:1993tx}. 
The entire physics community has largely endorsed its physics goals, since Japan, 
America and also Europe have set up many programs with similar purposes, and 
much progress has been achieved in 
the field. 

As already described in the original proposal, the final phase of the ICARUS project 
requires a sensitive mass of liquid Argon of 5000 tons or more. This is still true today, 
even after the advent of the SuperKamiokande with its fiducial mass of 
22.5 ktons~\cite{superkevid}.
The superior bubble-chamber-like features 
of the ICARUS detector will provide additional and fundamental contributions to the 
field.

Back in 1995, it was decided that the most conservative way to reach the liquid Argon 
sensitive mass of 5000 tons was to go through a
first step: the T600 detector~\cite{proposal2}.
This step-wise strategy allowed us to develop progressively the necessary know-how 
to build a large liquid Argon detector. 

As a yet additional step, a large $10m^3$
prototype was built in 1997, in order to assess the major 
issues concerning  cryogenics, internal detector mechanics and liquid Argon  
purification. The $10 m^3$ prototype has undergone several cooling and 
filling tests in Pavia; this phase successfully ended in July 1999. 
The $10m^3$ was then dismounted and transported to an external hall of
LNGS. Complementing the dewar with appropriate H.V., wire readout,
a scintillation light detection system and an external trigger, turned the
prototype into a fully functional liquid Argon imaging
chamber. A test, that lasted about 100~consecutive days, has allowed to prove
the technique in a configuration similar to the one adopted for the T600
detector and, thanks to a perfectly mastered LAr purification technique, 
has resulted in the collection of ionizing events of excellent
quality.

The T600 detector stands today as the first 
living proof that a large detector can be built and that liquid Argon
imaging technology can be 
implemented on such large scales. We are now ready to propose the
construction of a second T600 ``clone'', within a 24 months program, to complete
within 2003 the first 35~meters of the experimental hall, with 1.2~kton of
active liquid Argon mass.

After the successful completion of a series of technical tests to be
performed at the assembly hall in Pavia, the T600 detector will be ready to
be transported into the LNGS tunnel.

The operation of the T600 at the LNGS will allow us (1) to develop the local 
infrastructure needed to operate our large detector (2) to start the
handling
of the underground liquid argon technology
(3) to study the local background 
(4) to start the data taking with an initial liquid argon mass that
will reach in a 5-6 year program the multi-kton goal.
The T600 is to be considered as the first 
milestone on the road towards a total sensitive mass of 5000 tons: it is
the first piece of the detector to be complemented by further modules 
of appropriate size and dimensions,
in
order to reach in a most efficient and rapid way the final design mass.

In this document, we describe the physics program that will be
accomplishable with the first modules of liquid Argon. Given this initial
phase in which a limited amount of liquid Argon is available, we consider
the physics program achievable with exposures of 1 or 2 kton $\times$
year.

In section~\ref{dwmu}, we discuss the benchmark measurement provided by the
detection of downward-going muons. In section~\ref{sec:atm}, we address 
the detection
of atmospheric neutrinos, including fully-contained and partially-contained
events, and upward going muons. In section~\ref{sec:solar} the detection of
solar neutrinos is discussed. The sensitivity to nucleon decays is explored in
section~\ref{sec:nucdecay}. Finally, the detection of supernova neutrinos
is treated in section~\ref{sec:supernova}.











\section{Detection Of Downward--Going Muons}
\label{dwmu}

The rock overburden at Gran Sasso underground laboratory filters secondary
cosmic ray particles produced in atmospheric showers, and only high energy
muons (coming from the decay of secondary $\pi$ and K mesons) survive, with
a rate of about 1 particle /m$^2$ hour. The energy 
threshold for an atmospheric muon to reach the underground hall has an
exponential dependence on
the rock depth h($\theta$,$\phi$)~\cite{gaisser}, which is a function of
the direction, depending on the mountain topography.
In correspondence of the minimum thickness ($\sim$ 3100
hg/cm$^2$, in the direction of Campo Imperatore) $E_{thr}
\sim$ 1.3 TeV. The average residual energy of muons at the depth of Gran
Sasso underground hall is about 300 GeV. In their propagation through the
rock muons undergo different interaction processes affecting also their
direction. The average scattering angle has been calculated to be around
1 degree, mostly dominated by multiple scattering in the last part of their
path. 

The measurement of atmospheric muons surviving underground is not
a primary goal of ICARUS. However, it can be considered
a benchmark test of the detector performance, but also an interesting
by--product of the physics research program.
As a matter of fact, muons are practically the only available high energy 
particles with a quite constant rate which can be used to perform an effective
monitoring of the detector performance as far as track reconstruction is
concerned. An example of a possible
measurement is that of the muon flux as a function of 
the rock depth, or ``depth-intensity'' function $I(h)$. 
The comparison with the world average of $I(h)$~\cite{world}, and in particular with the
results 
from the previous experiments at Gran Sasso~\cite{lvd,macro}, provides a
check of the  
efficiency and stability of the detector operation.
Such a measurement requires a precise knowledge of the detector acceptance
and in addition, is a good check of the detector simulation codes.

From the physics point of view, the measurement of the underground muon flux
allows, by means of a detailed simulation of atmospheric showers, to 
extract the parameters of the all--nucleon flux of primary cosmic rays, in
the
region around a few TeV/nucleon, where also direct measurements suffer of
large errors. A better knowledge of the primary flux in this energy region is 
still important to reduce uncertainties in the calculations of the
flux of atmospheric neutrinos. The main interest is in a new analysis, 
more than in the statistics, which has been largely collected by MACRO and LVD,
since the interpretation is 
limited by both, theoretical systematics (hadronic interaction models) and
the knowledge of rock depth. At present, Gran Sasso
rock is known with an accuracy which is not better than a few percent.
Furthermore, further measurements at large zenith angle, {\it i.e.} at very
large values of h($\theta$,$\phi$), where the rate is quite low, are
important to extract informations 
about prompt muon production, connected to the charm production in
atmospheric showers~\cite{gaisser}. 
Also, the measurement close to the horizontal
direction is known to be dominated by atmospheric neutrino interaction in
the rock, and it is useful to establish the $\nu$ flux normalization in a
direction where the oscillation phenomena, according to the present
estimate of parameters, are marginal. 
In this large zenith angular region, the contribution of new data is in any
case important and ICARUS has at least a clear advantage, for
instance, with respect 
to MACRO, which did not present a specific analysis in this field.

\begin{figure}[bt]
\begin{center}
\mbox{\epsfig{file=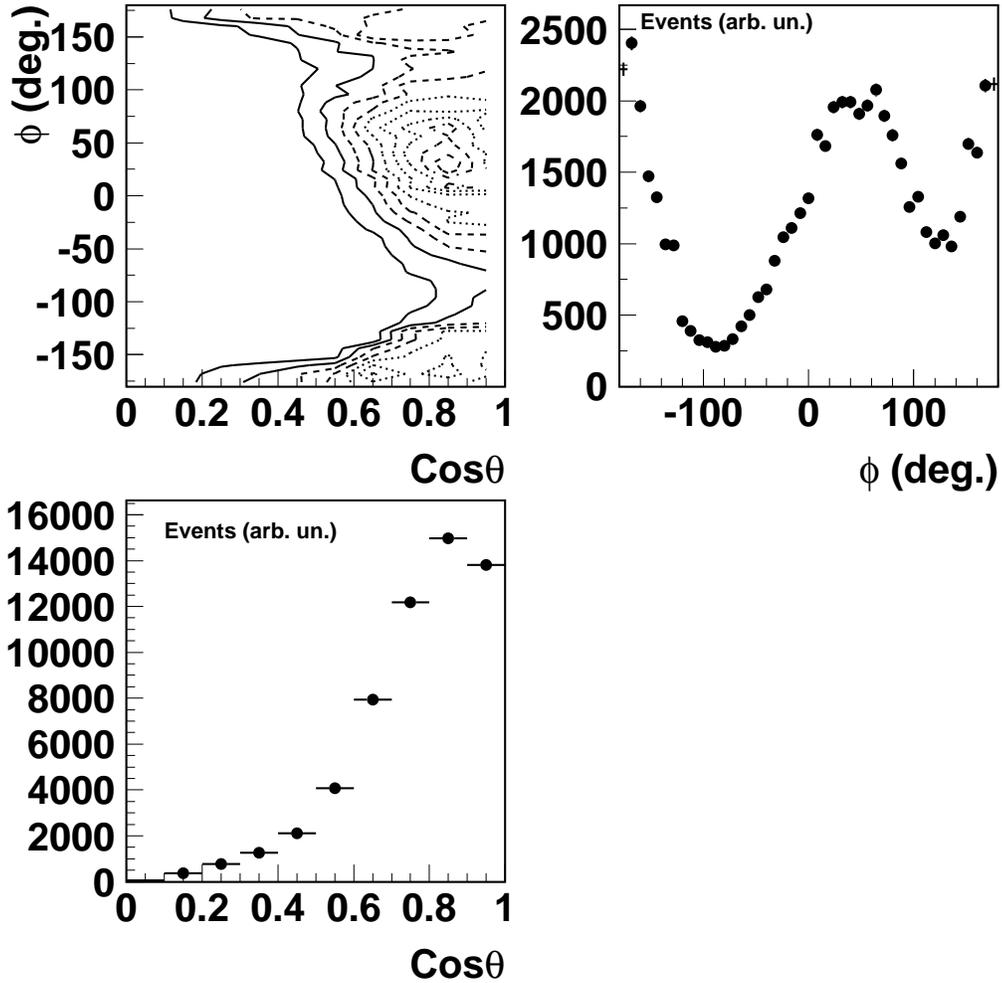,height=13cm}}
\caption{ Expected angular distribution of downward--going muons detected
in the T600 module.
\label{fig_mudw}}
\end{center}
\end{figure}

In order to evaluate the rate of downward--going muons detectable in the 
T600 module, we have performed a full simulation in the FLUKA 
environment~\cite{flukakek,flukalisb},
assuming that the module is located in the Hall B of the Gran Sasso
underground laboratory.
In order to generate the local muon flux, a $cos(zenith)$ - $\phi$ matrix,
unfolded from the MACRO experimental data, has been used.
For this first iteration of this kind of calculation, the contribution of
multiple muon events (about 6\% of 
the total muon event rate) has been neglected. To identify a muon,
we require a track of at least 20 cm, passing through at least 
one half--module.
The experience gained with the operation of the 10 m$^3$ module~\cite{10m3} allows us
to have confidence that, with this minimum track length, a 3--Dimensional
reconstruction is fully assured (at least 40--50 hits/view).
In this way we expect to detect 120 tracks/hour. The expected
angular distribution, the shape of which is dominated by the rock overburden, is
reported in Fig.\ref{fig_mudw}\footnote{There the $\phi$ angle is measured 
counter--clockwise from the axis parallel to the longitudinal section of
Hall B, which makes an angle of 128.4$^\circ$  with respect to the 
geographical North. The azimuth angle is normally defined as the angle 
measured clockwise from the North direction.}.

Within one year of live time, the data collected by the T600 module would
reach half the statistics collected by LVD in~\cite{lvd}.

From the above angular distribution the bin-by-bin muon intensity $ I(h)$
referred to 
the vertical direction is extracted as follows:
\begin{equation}
I(h)\ =\left ( {1\over{{\Delta}T}}\right
)
{{{\Sigma}_i}{N_i}{m_i}\over{\Sigma}{_j}{\Delta\Omega{_j}}A{_j}{\epsilon}{_j}/f{\theta}_j}
\end{equation}
\noindent where $\Delta$T is the live time; N$_i$ is the number of observed
events of muon multiplicity m$_i$
in the bin of slant depth h; A$_j$ is the effective detector projected area
for that bin;
${\epsilon}_j$ is the combined trigger and reconstruction efficiency and
$\theta_j$ is the muon zenith angle. The function $1/f{\theta_j}$
represents
the zenith distribution of muons in the atmosphere, which, for these energies and 
up to 60$^\circ$, is well
approximated by sec${\theta_j}$. For larger angles corrections must be
introduced to take into account the earth's curvature.
The projected area $A_{j}(\theta,\phi)$ and the detector tracking efficiency
$\epsilon_{j}(\theta,\phi)$ must be calculated from
a detailed Monte Carlo study. In order to compare experimental results from
different sites, it is customary to convert
the actual rock thickness to ``standard rock'' (Z=11, A=22) slant depth,
using the known chemical composition of  the Gran Sasso rock and a conversion
formula, like the one  described in~\cite{kotov}.

The topic of the study of atmospheric muons includes other items which can
be of interest for ICARUS.  In addition to the test runs foreseen in Pavia, 
the run of the T600 module in Gran Sasso is 
necessary to study the analysis techniques. Briefly we can mention the
following arguments:
\begin{enumerate}
\item Study of multiple muon events, which brings information about
the mass composition of primary cosmic rays. A fundamental advantage with
respect to the work performed by previous experiments, like
MACRO~\cite{macro_comp}, would be 
the possibility of reducing the error in the multiplicity determination
of the largest events, thanks to the fine grain spatial resolution.
In fact, the highest multiplicity events are found to be often composed 
of close tracks, concentrated in a relatively small area.

\item Study of the local energy spectrum of muons, exploiting the technique
of energy evaluation of muon tracks by means of the multiple scattering
evolution. 

\item Detailed study of radiative and photonuclear interactions of high
energy muons in the liquid Argon.

\end{enumerate}


\section{Atmospheric Neutrinos}
\label{sec:atm}

\subsection{Contained and partially contained events}
The comprehensive investigation of atmospheric neutrino events, beyond what
was already achieved in SuperKamiokande, requires a fiducial mass of several
ktons, in order to reach the level of at least one thousand events per
year. The physics goals of new atmospheric neutrino measurements are to firmly
establish the phenomenon of neutrino oscillations with a different
experimental
technique, possibly free of systematic biases, measure the oscillation
parameters and clarify the nature of the oscillation mechanism.

The statistics accumulated in a 2 kton $\times$ year exposure
will be modest (see Table~\ref{tab:atm_2kton}). They will be comparable
to those obtained in the first generation of water Cerenkov detectors,
namely Kamiokande and IMB.

\begin{table}[htbp]
\begin{center}
\begin{tabular}{lccccc}
 &\multicolumn{5}{c}{2 kton$\times$year}\\\cline{2-6}
 & & \multicolumn{4}{c}{$\Delta m_{23}^2$ (eV$^2$)} \\\cline{3-6}
 & No osci & $5\times 10^{-4}$ & $1\times 10^{-3}$ & $3.5\times 10^{-3}$ & 
$5\times 10^{-3}$ \\ \hline
\multicolumn{6}{c}{}\\
{\bf Muon-like} & $270\pm16$ & $206\pm14$ & $198\pm14$ & $188\pm14$ & $182\pm13$ \\
\multicolumn{6}{c}{}\\
\hspace*{1.cm}Contained & $134\pm12$ & $100\pm10$ & $96\pm10$ & $88\pm9$ & $86\pm9$ \\
\hspace*{1.cm}Partially-Contained & $136\pm12$ & $106\pm10$ & $102\pm10$ & $100\pm10$ & $96\pm10$ \\
\multicolumn{6}{c}{}\\
\hspace*{1.cm}No proton & $104\pm10$ & $76\pm9$ & $74\pm9$ & $68\pm8$ & $66\pm8$ \\
\hspace*{1.cm}One proton & $82\pm9$ & $64\pm8$ & $60\pm8$ & $58\pm8$ & $56\pm7$ \\
\hspace*{1.cm}Multi-prong & $84\pm9$ & $66\pm8$ & $64\pm8$ & $62\pm8$ & $60\pm8$ \\
\multicolumn{6}{c}{}\\
\hspace*{1.cm}$P_{lepton}<400$ MeV & $114\pm11$ & $82\pm9$ & $80\pm9$ & $74\pm9$ & $70\pm8$ \\
\hspace*{1.cm}$P_{lepton}\geq400$ MeV & $156\pm12$ & $124\pm11$ & $118\pm11$ & $114\pm11$ & $112\pm11$ \\ \hline
\multicolumn{6}{c}{}\\
{\bf Electron-like} & $152\pm12$ & $152\pm12$ & $152\pm12$ & $152\pm12$ &
$152\pm12$ \\
\multicolumn{6}{c}{}\\
\hspace*{1.cm}Contained & $100\pm10$ & $100\pm10$ & $100\pm10$ & $100\pm10$ & $100\pm10$ \\
\hspace*{1.cm}Partially-Contained & $52\pm7$ & $52\pm7$ & $52\pm7$ & $52\pm7$ & $52\pm7$ \\
\multicolumn{6}{c}{}\\
\hspace*{1.cm}No proton &$64\pm8$ &$64\pm8$ &$64\pm8$ &$64\pm8$ &$64\pm8$ \\
\hspace*{1.cm}One proton & $48\pm7$ & $48\pm7$ & $48\pm7$ & $48\pm7$ & $48\pm7$ \\
\hspace*{1.cm}Multi-prong & $40\pm6$ & $40\pm6$ & $40\pm6$ & $40\pm6$ & $40\pm6$ \\
\multicolumn{6}{c}{}\\
\hspace*{1.cm}$P_{lepton}<400$ MeV & $74\pm9$ & $74\pm9$ & $74\pm9$ & $74\pm9$ & $74\pm9$ \\
\hspace*{1.cm}$P_{lepton}\geq400$ MeV & $78\pm9$ & $78\pm9$ & $78\pm9$ & $78\pm9$ & $78\pm9$ \\\hline
\multicolumn{6}{c}{}\\
{\bf NC-like} & $192\pm14$ & $192\pm14$ & $192\pm14$ & $192\pm14$ & $192\pm14$ \\\hline
\multicolumn{6}{c}{}\\
{\bf TOTAL} & $614\pm25$ \\\hline

\end{tabular}
\end{center}
\caption{Expected atmospheric neutrino rates in case no oscillations 
occur and assuming $\nu_\mu\to\nu_\tau$ oscillations take place with
maximal mixing. Four different $\Delta m^2$ values have been
considered. Only statistical errors are quoted.}
\label{tab:atm_2kton}
\end{table}

The capability to observe all processes, electron, muon and tau
neutrino charged current events (CC) and all neutral currents (NC) without
detector biases and down to kinematical threshold, 
will however provide an unique new view on the atmospheric events.

The ICARUS T600 will offer an observation
of atmospheric neutrinos of a very high quality, 
thanks to
its unique performance in terms of resolution and precision.

The perspective of ICARUS is
to provide redundant, high precision measurement and
minimize as much as possible the systematics uncertainties
of experimental origin which affect the results of existing experiments.
Improvements over existing methods are expected in
\begin{enumerate}
\item neutrino event selection
\item identification of $\nu_\mu$, $\nu_e$ and $\nutau$ flavors
\item identification of neutral currents
\end{enumerate}
The operation of the T600 will be the only way to demonstrate
{\it in situ} the expected performance of the liquid Argon technique.

Unlike measurements obtained up to now in Water Cerenkov detectors,
which are in practice limited to the analysis of ``single-ring'' events,
complicated final states with multi-pion products, occurring mostly
at energies higher than a few GeV, will be completely analyzed and
reconstructed in the T600. This will be a significant improvement
with respect to previous observations. 
As an example, we anticipate a much better angular 
resolution of the reconstructed direction of the incoming neutrino. 
The reconstruction of the zenith angle of the incoming $\nu$ is of
great importance in the search of oscillations in atmospheric
neutrinos. In SuperKamiokande measurements,
the direction of the incoming neutrino is taken to be the
one of the leading lepton, since due to pattern recognition, 
only single ``ring'' events are analyzed. ICARUS allows
for a better
reconstruction of the incoming neutrino variables (i.e. incidence
angle, energy) by using the information coming from all particles
produced in the final state.

\begin{figure}[htbp]
\begin{center}
\mbox{
\epsfig{file=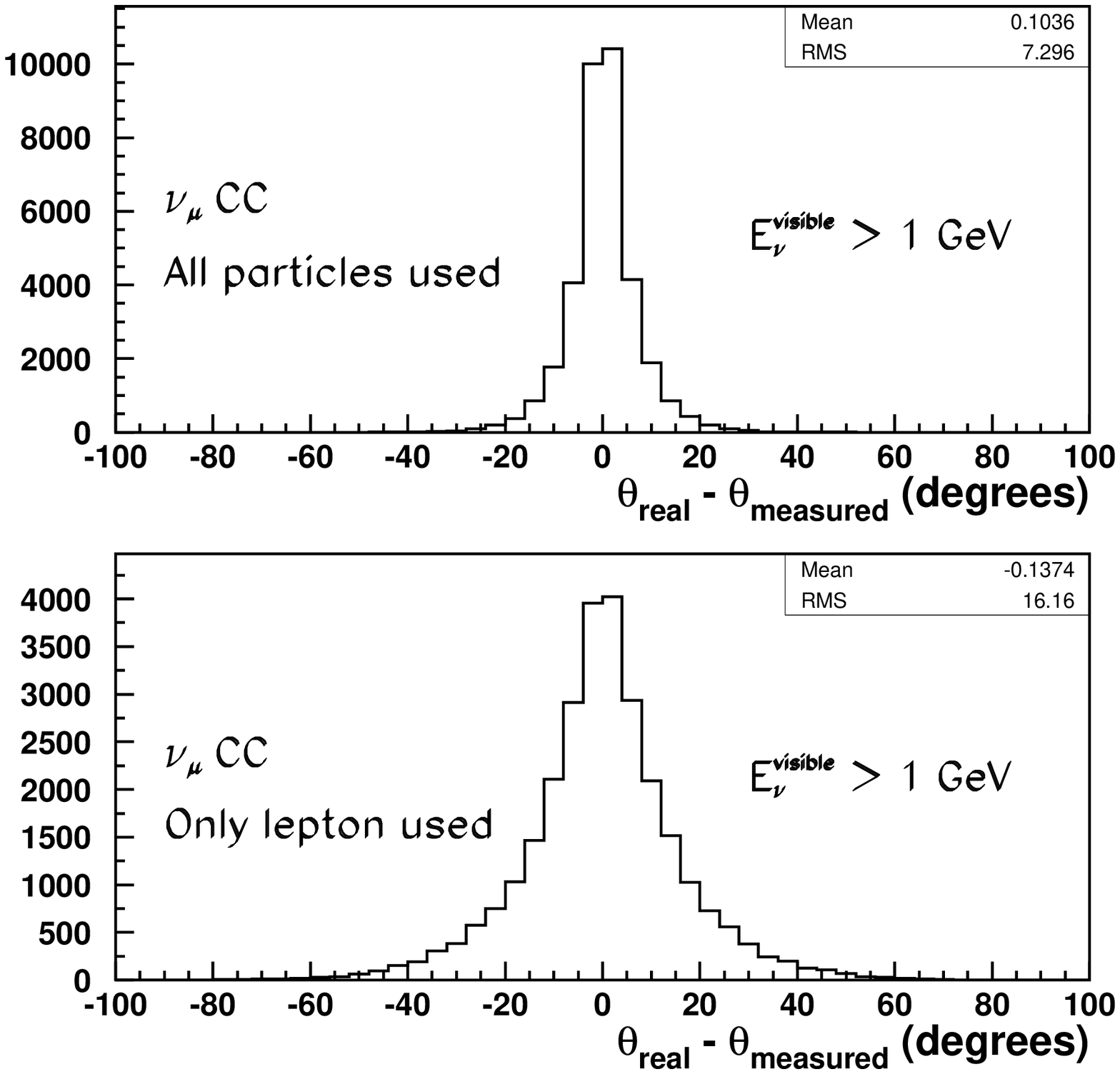,width=8cm}
\epsfig{file=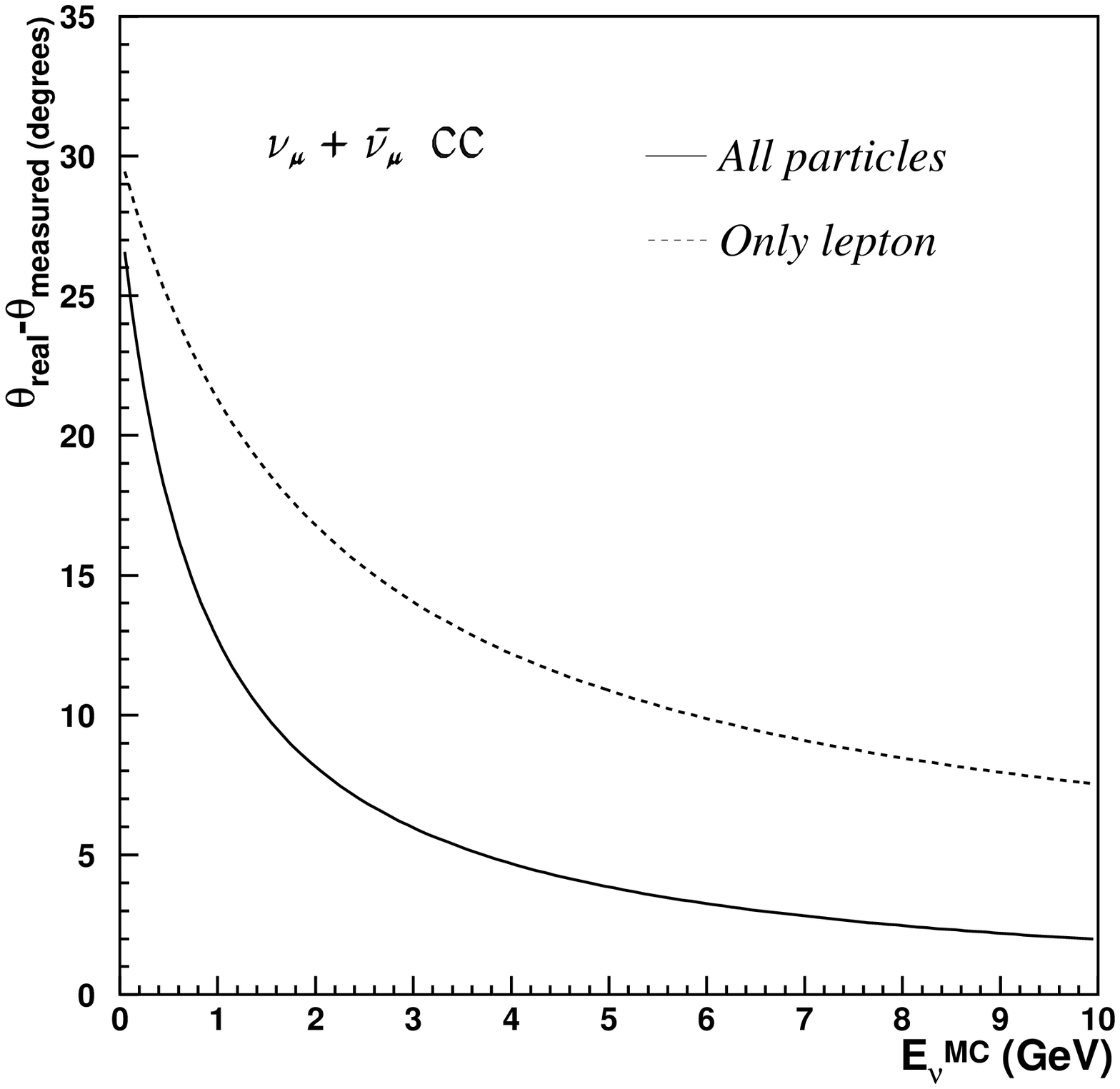,width=8cm}
}
\end{center}
\caption{(left) Zenith angle resolution. The top plot shows the resolution
obtained by reconstructing the incoming neutrino direction using all
particles momenta, the bottom plot shows the resolution obtained using
only the leading lepton momentum.
(right) Zenith angle resolution as a function of the neutrino energy.}
\label{ZEN-RES}
\end{figure}

Figure~\ref{ZEN-RES}(left) shows the distribution of the difference 
between the real and reconstructed neutrino angle 
for the whole sample of events with $E_{\nu}>1$~GeV.  
The improvement on the angular resolution is visible.
The RMS of the distribution improves from $\sim 16$ to $\sim 8$ degrees
after the inclusion of the hadronic jet in the reconstruction.

Figure~\ref{ZEN-RES}(right) shows the zenith angle resolution as a
function of the incoming neutrino energy, comparing the
two methods of reconstruction. For energies tending to zero, the resolution
is dominated by the smearing introduced by the Fermi motion of the
initial state nucleon and by re-interaction of the hadrons inside
the nucleus, and therefore the improvement obtained with the hadronic
jet is minimal. For energies above $\approx 500\ \rm MeV$, the improvement
in resolution is significant. It should be noted that at higher energies, 
the inclusion of the hadronic jet improves by a factor three 
the reconstruction of incoming neutrino direction. This will allow to have 
a more precise reconstruction of the neutrino L/E (see ICANOE proposal~\cite{icanoe}). 

We recall the expected atmospheric neutrino rates
obtained per year for an exposure of 2 ktons in Table~\ref{tab:atm_2kton}, 
with and without $\numu\ra\nutau$ oscillation hypothesis ($\sin^22\theta=1$).

Muon-like events contain an identified muon and correspond
to $\numu$/$\bar\numu$ CC events. Electron-like
are events with an identified electron and are 
$\nue$/$\bar\nue$ CC events.
Given the clean event reconstruction, the ratio $R$ of ``muon-like''
to ``electron-like'' events can be determined free of large experimental
systematic errors. In fact, the expected purity of the samples
is above 99\%. In particular, the contamination from $\pi^0$
in the ``electron-like'' sample is expected to be completely
negligible.

\begin{figure}[htbp]
\centering
\epsfig{file=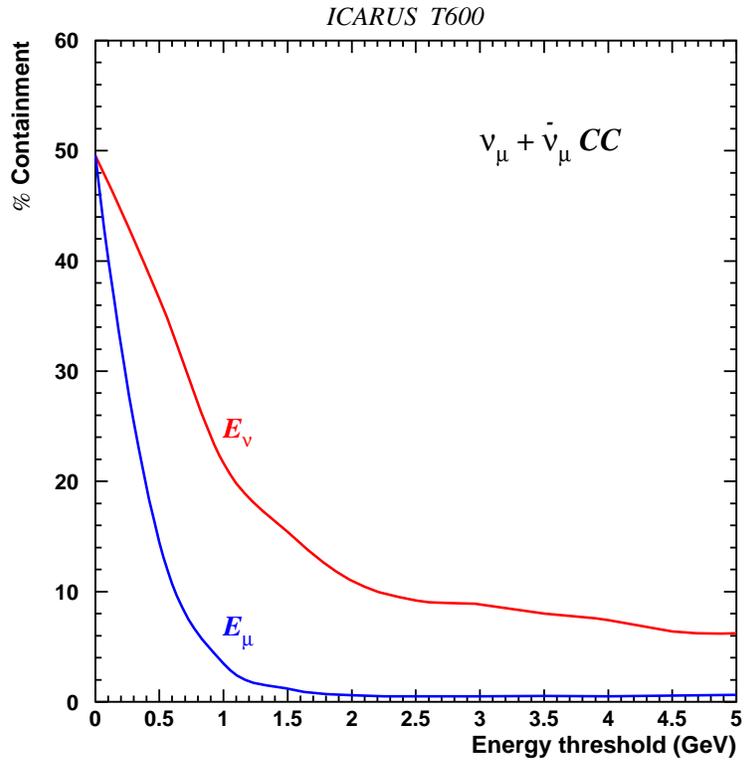,width=11cm}
\epsfig{file=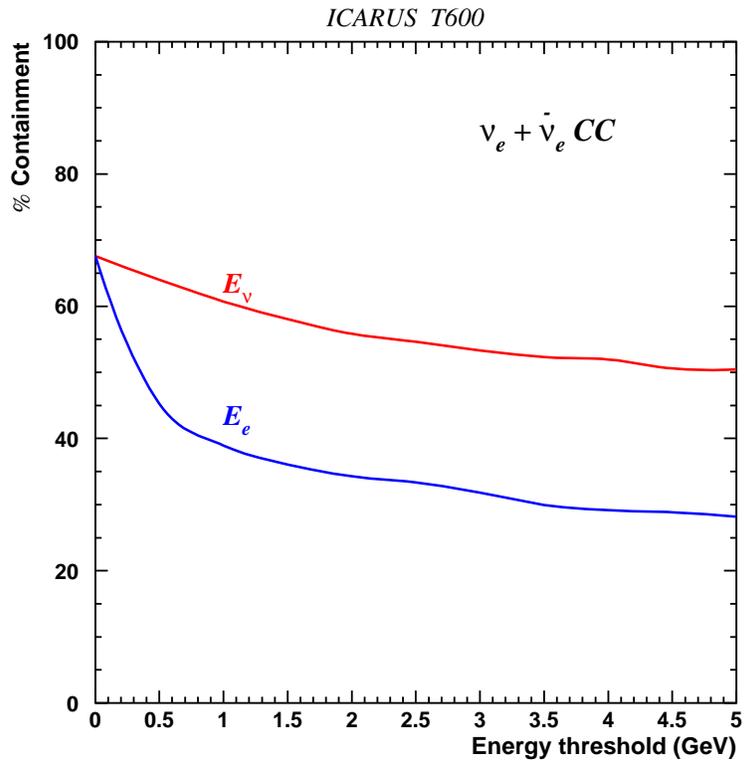,width=11cm}
\caption{Integral distributions showing the containment for CC events as a function of the neutrino 
energy and the leading lepton momentum.Top plot: $\nu_\mu + \bar\nu_\mu$ CC. Bottom plot: 
$\nu_e + \bar\nu_e$ CC.}
\label{fig:t600cont}
\end{figure}

We further split the muon and electron-like events into fully-contained and
partially-contained
samples. ``Fully contained events'' are those for which the visible products
of the neutrino interaction are completely contained within
the detector volume. 
``Partially contained events'' are events
for which the leading lepton exits the detector volume.
Figure~\ref{fig:t600cont} shows the containment of charged
current events for different incoming
neutrino energies and leading lepton momentum thresholds. The top plot refers 
to $\nu_\mu + \bar\nu_\mu$ CC events and the bottom plot 
to $\nu_e + \bar\nu_e$ CC.
In the computation, events were generated uniformly over the full T600 
active argon volume and were oriented
correctly in the detector volume. Particles were tracked
through the argon until they reached the wall of the argon
volume. 

Muon-like events are less contained than electron-like ones.
The overall fraction of contained events
is $50\%$ for $\nu_{\mu} + \bar{\nu}_{\mu}$ CC events. 
This fraction decreases rapidly with muon momentum. For
muon momenta above $2\rm\ GeV$, in practice, all muons
will escape the mother volume. Since at low energies,
muons carry on the average
more than half of the incoming neutrino energy, this strong
dependence of the containment is also visible as a function
of the incoming neutrino energy (see Figure~\ref{fig:t600cont}, top plot).
The fraction of contained $\nu_e + \bar{\nu}_e$ CC is close to $70\%$ of 
the total expected rate. The decrease of containment with energy is not so 
dramatic for electron-like events. For neutrino energies in excess of 5 GeV, 
we expect half of the $\nu_e + \bar{\nu}_e$ CC sample to be fully contained 
(see Figure~\ref{fig:t600cont}, bottom plot).

For the muon-like contained events, the muon energy is
precisely determined by integration of the $dE/dx$ measurements
along the track. For partially contained events, in which the muon
escapes the detector active volume, the muon momentum is estimated
via the multiple scattering method (see ICANOE proposal~\cite{icanoe}).
Fully contained electromagnetic showers are extremely well measured 
thanks to the superb calorimetric performance of liquid Argon.
The energy of partially contained showers can be recovered by a careful 
``shower-shape'' analysis.

We also illustrate the expected event rates classified according
to their final state multiplicity (see Table~\ref{tab:atm_2kton}). 
Approximately 40\% of CC events
contain no proton in the final state\footnote{A proton
is identified if its kinetic energy is above 50~MeV.}, corresponding
to the ``one-ring'' sample. The rest of the events will contain
a proton or multi-prongs final states, which will provide, thanks
to the precise reconstruction of all particles, a precise determination
of the incoming neutrino energy and direction.

Finally, we also point out that in ICARUS atmospheric neutrino
events can be analyzed down to production threshold, given the
excellent imaging. We illustrate this by classifying
the events according to the energy of
the leading lepton (electron or muon). 
We split the samples into
$P_{lepton} < 400$ MeV and $P_{lepton} > 400$ MeV, which 
correspond to the threshold used in Super-Kamiokande~\cite{superkevid}.
Almost $50\%$ of the expected rate lies below the threshold
and hence ICARUS can really contribute to the understanding
of the low energy part of the atmospheric neutrino spectrum.

An improved observation of about 100 neutral current (NC) events
per kton $\times$ year is also
expected, given the clean classification of events based
on the absence of an electron or muon in the final state.
In this case, the excellent $e/\pi^0$ separation plays a
fundamental role to select an unbiased, free of background
neutral current sample.

\begin{table}[ht]
\begin{center}
\begin{tabular}{lccccc}
 &\multicolumn{5}{c}{2 kton$\times$year}\\\cline{2-6}
 & & \multicolumn{4}{c}{$\Delta m_{23}^2$ (eV$^2$)} \\\cline{3-6}
 & No osci & $5\times 10^{-4}$ & $1\times 10^{-3}$ & $3.5\times 10^{-3}$ & 
$5\times 10^{-3}$ \\ \hline
\multicolumn{6}{c}{}\\
{\bf Muon-like} & $270\pm16$ & $206\pm14$ & $198\pm14$ & $188\pm14$ & $182\pm13$ \\
\multicolumn{6}{c}{}\\
\hspace*{1.cm}Downward & $102\pm10$  & $102\pm10$   & $102\pm10$  & $98\pm10$  & $95\pm10$ \\
\hspace*{1.cm}Upward & $94\pm10$  & $46\pm7$  & $46\pm7$  & $47\pm7$  & $49\pm7$ \\\hline
\multicolumn{6}{c}{}\\
{\bf Electron-like} & $152\pm12$ & $152\pm12$ & $152\pm12$ & $152\pm12$ &
$152\pm12$ \\
\multicolumn{6}{c}{}\\
\hspace*{1.cm}Downward & $56\pm7$  & $56\pm7$  & $56\pm7$  & $56\pm7$  & $56\pm7$ \\
\hspace*{1.cm}Upward & $48\pm7$  & $48\pm7$  & $48\pm7$  & $48\pm7$  & $48\pm7$ \\\hline

\end{tabular}
\end{center}
\caption{Predicted downward ($\cos{\theta_{zenith}}> 0.2$) and 
upward ($\cos{\theta_{zenith}}< -0.2$) atmospheric neutrino rates in 
case no oscillations occur and assuming $\nu_\mu\to\nu_\tau$ oscillations 
take place with maximal mixing. Four different $\Delta m^2$ values have been
considered. Only statistical errors are quoted. As a reference, we also show 
the total expected rates for both muon and electron-like events.}
\label{tab:atm_2kton_up_down}
\end{table}

The presence of neutrino oscillations leads to differences in the
predicted rates of upward and downward going neutrino events (see 
Table~\ref{tab:atm_2kton_up_down}). For a 2 kton $\times$ year
exposure, we will measure a quite evident deficit of upward-going 
muon-like events, for the range of oscillation parameters presently allowed 
by Super-Kamiokande measurements.

The question whether the atmospheric neutrino anomaly is due to 
$\nu_\mu \to \nu_\tau$ or $\nu_\mu \to \nu_s$ 
oscillations is not totally settled~\cite{Fukuda:2000np}. The clean NC sample will allow us to perform 
an ``indirect'' $\nu_\tau$ appearance search. To discriminate between $\nu_{\mu} \to \nu_{\tau}$ 
and $\nu_{\mu} \to
\nu_s$ oscillations, we measure the ratio $R_{NC/e} =
\frac{NC^{obs}/\nu_e CC^{obs}}{NC^{exp}/\nu_e CC^{exp}}$. An
oscillation to an active neutrino leads to $R_{NC/e} =1$, while
$R_{NC/e} \sim 0.7$ is expected for an oscillation to a sterile
neutrino.

The value and error of $R_{NC/e}$, in case of oscillation to active
neutrino, are shown in Table~\ref{tab:rnce}, either using all events 
or only fully contained ones. The systematic error is expected to be low, since our 
measurement of the double ratio does not depend on poorly known cross sections 
(e.g., single $\pi^0$ production). The expected error on $R_{NC/e}$ is $22\%$ ($15\%$) 
for an exposure of 1 (2) kton $\times$ year. The quoted uncertainty is similar to the 
one obtained by Super-Kamiokande (which is strongly dominated by systematics).

\begin{table}[htbp]
\begin{center}
\begin{tabular}{|c|c|c|}
\hline
Exposure (kton$\times$year) & \multicolumn{2}{c|}{$R_{NC/e}$} \\
& all events & contained \\
\hline
 1 & $ 1.0 \pm 0.22$ & $ 1.0 \pm 0.23$ \\
\hline
2 & $ 1.0 \pm 0.15$  & $ 1.0 \pm 0.16$ \\
\hline
5 & $ 1.0 \pm 0.10$  & $ 1.0 \pm 0.10$ \\
\hline
\end{tabular}
\end{center}
\caption{$R_{NC/e}$ as a function of the exposure assuming oscillation 
to an active neutrino. Quoted errors are of statistical nature.}
\label{tab:rnce}
\end{table}

\subsection{Upward--Going Muons}

Muon neutrinos undergoing CC interactions in the rock surrounding the
detector can produce muon
tracks detectable by the apparatus. 
These events are produced
by neutrinos belonging to an energy region which is
higher with respect to that of contained or partially contained
events: typically they cover the range from Multi--GeV up to
few TeVs, with a maximum in the region around 100 GeV. 
Therefore they are an interesting complement 
to the analysis of atmospheric
neutrinos, since the disappearance due to the oscillation process 
(or other disappearance mechanisms) will be quantitatively different with
respect to contained, or partially contained events. 
Furthermore, these events are also affected
by different systematics, not only because they belong to a different
region of the spectrum, but also because of different cross
sections and target material (the rock). 

Upward--going muons allow to explore path lengths through the
Earth ranging from few hundred km up to $\sim$ 13000 km. 
Neutrino oscillation will modify the total number of events and their
zenith angle distribution, mostly around the vertical direction, 
as already discussed in the work of
Super--Kamiokande~\cite{skthrou} and MACRO~\cite{upmumacro}.
Furthermore, the analysis of the angular distribution of
upward--going muons is one of the tools to discriminate 
between $\nu_\mu$--$\nu_s$ and $\nu_\mu$--$\nu_\tau$ 
oscillations~\cite{Lipari:1998rf}.


It is clear that the T600 module
is not large enough to provide statistically competitive results with
respect to MACRO or Super--Kamiokande. 
Instead, its operation at Gran Sasso allows to test 
a specific capability of ICARUS to recognize track direction,
thanks to the fine spatial resolution, which allows to identify
and reconstruct
$\delta$--rays produced along the track.

If an energy cut is applied in
order to select sufficiently energetic $\delta$ electrons, 
then these are expected to be emitted in the direction of the parent.
In fact, the cosine of the angle $\Theta$ between the $\delta$ and the
primary muon, 
for energies much larger than the atomic levels,
is given by the following expression:
\begin{equation}
\cos\Theta = \frac{T_e}{P_e} \frac{(E_\mu - m_e)}{P_\mu}
\end{equation}
where $T_e$ and $P_e$ are the kinetic energy and the momentum of the
emitted electron, $E_\mu$ and $P_\mu$ are the total energy and the momentum
of the parent muon and $m_e$ is the electron mass.

Therefore, practically independently of $E_\mu$, $\delta$ electrons of a few
MeV are emitted at small angle. Requiring a threshold of 10 MeV, about 
one $\delta$ ray of this kind is produced for each meter of track.
An electron of 10 MeV has a range slightly exceeding 5 cm, and,
with the wire pitch of 3 mm, its track can be sampled 15 times. 
The multiple
scattering in argon  will be relevant only at the end of $\delta$ electron range, while in the
first part of the track, the multiple scattering $\theta^{rms}~\sim$ 5$^\circ$ in one cm.
In order to have a good rejection power, it is therefore necessary to
ask for a minimum track length of 2$\div$3 m, with at least 2 $\delta$'s
above threshold. 
Monte Carlo simulations show that this method
allows to achieve the required rejection power.
In the case of low energy muons, a redundant method to identify the
direction can come from the analysis of the evolution of the multiple
scattering angle along the track.

\subsubsection{Experimental test with the 50 liter prototype}

In order to perform a test of the actual reconstruction ability of the $\delta$-rays direction 
on real muon tracks, we took advantage of the data taken with the ICARUS 50 liter LAr 
TPC exposed at the CERN neutrino beam during the 1997 physics run. The detector 
and the experimental set-up has been extensively described elsewhere~\cite{50L}. 
The results discussed here come from a recent re-analysis of these data~\cite{delta}, 
which took advantage of the current developments on the reconstruction program.



The collected muon event sample (about one thousand) was used to study the possibility 
to exploit the visible $\delta$-rays to determine the direction of minimum ionizing particle 
tracks crossing the detector. It allowed us to verify on real data the 
minimum energy allowing full reconstruction of the $\delta$-ray direction and, as a 
consequence, the effective rate that one can finally expect.



The analysis proceeded through the following steps.
The whole sample of through-going minimum ionizing particle tracks was visually 
scanned. Only the events containing a muon fully reconstructed in NOMAD and 
matching a track in the LAr-TPC were retained. This requirement safely predefined the 
direction of the selected tracks in the LAr-TPC.

\begin{figure}[htpb]
\centering
\epsfig{file=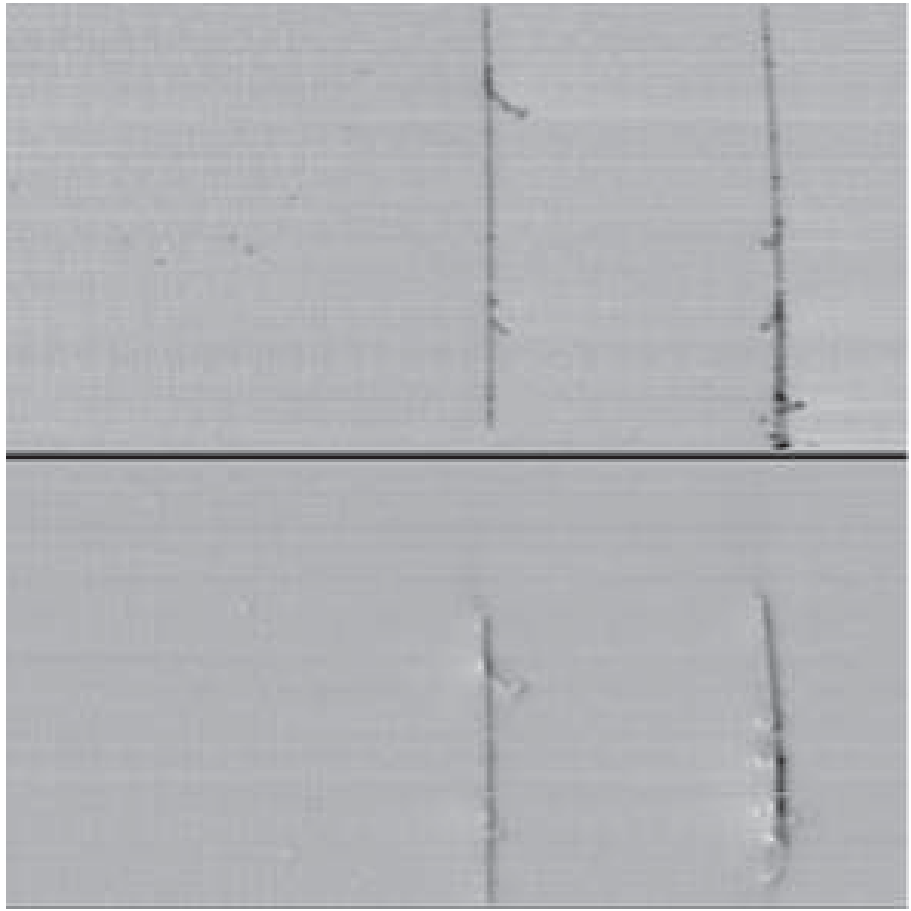,width=12cm}
\caption{An example of muon track recorded in the ICARUS 50 liter liquid Argon TPC prototype 
exposed at the CERN neutrino beam. The horizontal axis is the drift time; the vertical one is the wire 
numbering (top is the collection view, bottom is the induction plan). The visible area in each view 
corresponds to $472 \times 325$ mm$^2$. The muon enters from the top of the
picture on both views. Some 
$\delta$-rays 
are clearly visible. The track at the extreme right of 
the picture is not matched by NOMAD and exhibits a showering electron.}
\label{fig:drays1}
\end{figure}



The selected muon sample consists of 
320 events, corresponding to 120 m of minimum ionizing particle track.
The search and reconstruction of $\delta$-rays were then performed following a series of 
simple criteria.  A visible $\delta$-ray was defined as a track starting as double ionization on top of the 
main muon track and stopping aside of the main track in at least one view.

The $\delta$-ray kinetic energy was simply calculated from the reconstructed 3-D range 
of its track, knowing that the average $dE/dx$ is about $2.1$ MeV/cm practically 
constant over all the range.  Only $\delta$-rays with deposited energy above 2 MeV and below 
30 MeV (critical energy in LAr) were retained.


The above criteria selected 235 $\delta$-rays over the total 120 meters of tracks, 
namely about 2 $\delta$-rays per meter, with kinetic energy larger then 2 MeV. Remarkably, none was identified with the wrong direction. 


In order to understand our results, we compared the experimental $\delta$-rays energy 
spectrum with the predicted rate, valid for kinetic a energy much higher 
than the mean 
excitation energy ($T \gg 188 \rm\ eV$ for Argon)~\cite{PDG}:
\begin{eqnarray}
\frac{d^2N}{dTdx} &\approx& \frac{1}{2}K\rho \frac{Z}{A}\frac{1}{\beta^2}
\frac{1}{T^2}F_1F_2\\
&=&\frac{9.67F_1F_2}{(T/MeV)^2}m^{-1}MeV^{-1}
\label{eq:deltaequ}
\end{eqnarray}

To account for the selection requirements two factors, $F_1$, $F_2$, were included in 
Equation~\ref{eq:deltaequ}. The containment factor, F1, was used to account for the fact that not all the 
muon track length was available as origin of a $\delta$-ray because a fraction of it was needed 
to contain the $\delta$-ray. $F_1$ decreases with increasing $\delta$-ray energy.

The fraction of events, the end-point of which was separated from the muon track by 
more than 6 mm in at least one of the two 2-D views, was evaluated by means of the 
multiple scattering formula. The separation factor, $F_2$, was thus computed as a function 
of the kinetic energy of the $\delta$-rays. $F_2$ increases with increasing
$\delta$-ray energy.

\begin{figure}[htbp]
\centering
\epsfig{file=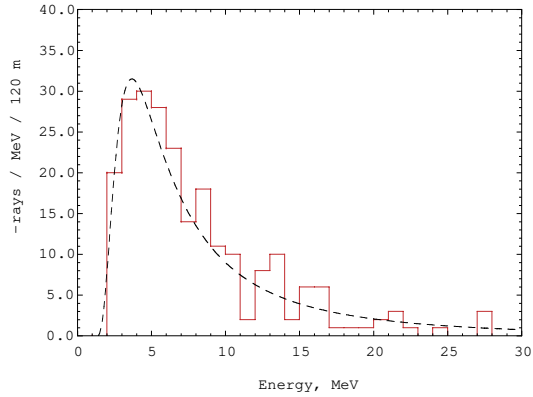,width=7cm}
\caption{Energy distribution of the $\delta$-rays: 
experimental data (solid histogram) are plotted together with 
the expected rate normalized to 120 m of m.i.p. track (dashed curve).}
\label{fig:drays2}
\end{figure}

\begin{figure}[htpb]
\centering
\epsfig{file=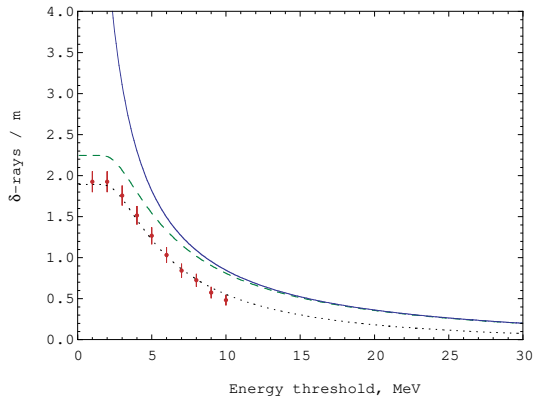,width=7cm}
\caption{Experimental cumulative energy distribution of the $\delta$-rays
(solid circles). 
The expectations are 
also plotted: the solid curve gives the rate with no selection cuts, the dotted line takes into account the 
factors $F_1$ and $F_2$ described in the text, the dashed one includes only $F_2$.}
\label{fig:drays3}
\end{figure}

Figure~\ref{fig:drays2} shows the energy spectrum of the experimental data of this test 
superimposed to that predicted with Equation~\ref{eq:deltaequ} normalized to 120 meters of track. 
Note, that the maximum rate occurs for a kinetic energy of 4 MeV 
and it decreases rapidly at low energy side because of the 
separation requirement.

Figure~\ref{fig:drays3} shows the integrated $\delta$-ray spectrum as a function of the energy threshold 
normalized to one meter of track. Data are plotted together with the predictions. The 
solid curve gives the rate with no cuts. The dotted line instead takes into account the 
factors $F_1$ and $F_2$: the agreement with the data is satisfactory.

The dashed line, which included only the factor $F_2$, has also been plotted because it 
gives the rate expected in a large LAr-TPC where the $\delta$-rays containment factor, $F_1$, is 
close to 100 \%. About 2.25 reconstructed $\delta$-rays per meter of track are predicted with 
energy above 2 MeV.

We demonstrated experimentally that in the ICARUS LAr-TPC we could make use 
of the $\delta$-rays kinematics to determine the direction of a minimum ionizing particle track 
crossing the detector~\cite{delta}. Slightly more than two $\delta$-rays per meter of track should be fully 
reconstructed. This implies that the ability to reconstruct 
track direction is very high: an 
efficiency of 99 \% is at reach considering only 2 m of track.

\subsubsection{Expected event rate at LNGS}

We have considered a full simulation in the FLUKA environment~\cite{flukakek,flukalisb} of the 
600 tons module of ICARUS.
Upward--going muons have been generated with the local expected
spectrum and angular distribution predicted by the Bartol
model~\cite{agrawal}
and considering a statistics equivalent to 30 years of operation.
A minimum track length of 2 meters has been requested.
In the no-oscillation case, 37 events/year are expected in the T600 module.
The additional requirement of producing at least 2 $\delta$'s of energy greater
or equal to 10 MeV reduces this number to 29 useful events. 

\begin{table}[htbp]
\begin{center}
\begin{tabular}{|c|c|}
\hline
$\Delta$m$^2$ & Events/year  \\
(10$^{-3}$ eV$^{2}$) &      \\ 
\hline 
1 & 23 \\
3 & 20 \\
5 & 13 \\
\hline
\end{tabular}
\caption{Expected rate of reconstructed upward going muon events
in the T600 module for a
minimum track length of 2 m and asking for the identification of at least
 2 $\delta$s with E$\ge$ 10 MeV.\label{tb:upmu1}}
\end{center}
\end{table}

\begin{figure}[htpb ]
\begin{center}
\mbox{\epsfig{file=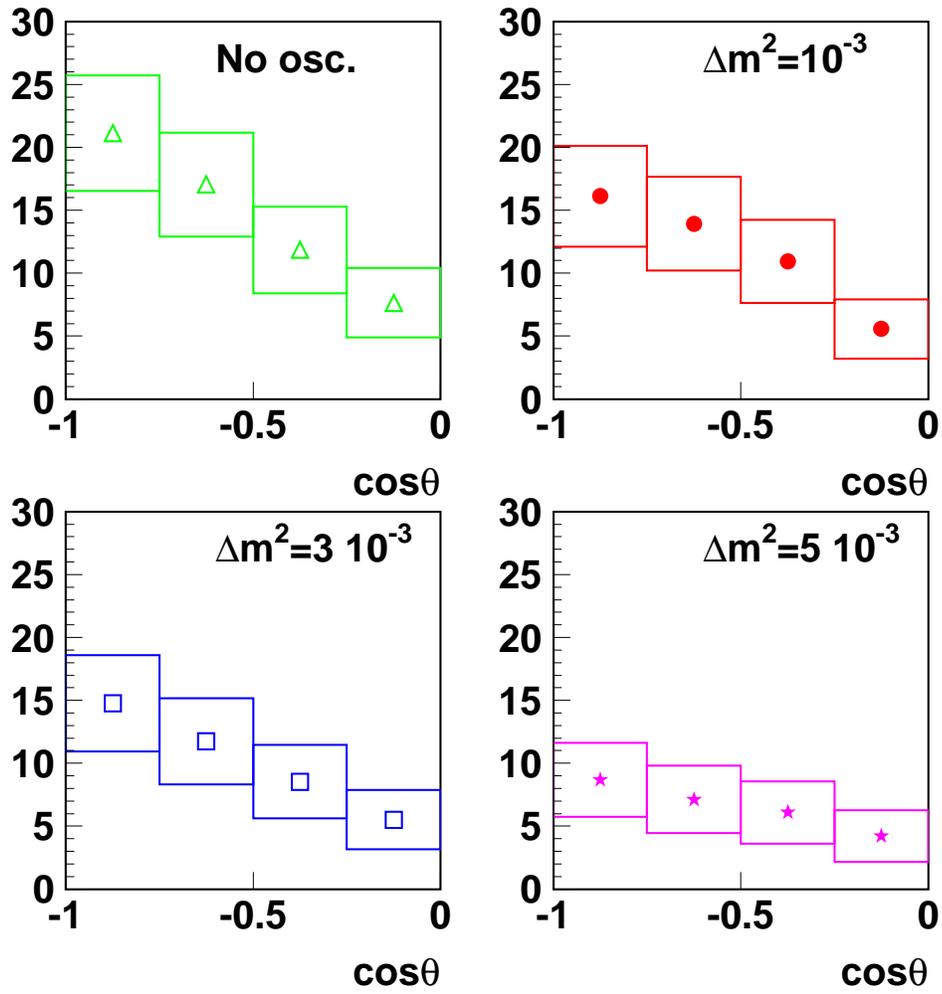,height=13cm}}
\caption{
Expected zenith angle distributions achievable in 2 years (live time)
of operation for the different values of
$\Delta$m$^2$ (sin$^2 2\theta$=1)
as a function of the cosine
of the zenith angle. The box shows the statistical error. 
\label{fig_upmu}}
\end{center}
\end{figure}

Table \ref{tb:upmu1} summarizes these results for different values
of the $\Delta$m$^2$ oscillation parameter in the maximal mixing
$\nu_\mu$--$\nu_\tau$ scenario.

The expected zenith angle distributions achievable in 2 years (live time)
of operation for the different values of
$\Delta$m$^2$ are reported in Fig. \ref{fig_upmu}
 
In addition, we can mention that the particle identification capability in
the liquid Argon 
allows to have further rejection of the background due to upward--going 
charged hadrons locally produced in the interactions of 
downward--going muons~\cite{hadron}.



\section{Solar Neutrinos}
\label{sec:solar}

The unique capabilities of a liquid Argon TPC are suitable to
the real-time detection of the neutrinos produced in the Sun. 
Two independent neutrino reactions contribute to the total expected rate: 
elastic scattering by electrons and absorption on Argon nuclei. These 
interactions usually result in the production of a 
primary electron track, sometimes accompanied by secondary 
electron tracks of lower energy.  
 
Small liquid Argon TPC prototypes have demonstrated that 
electrons with a kinetic energy as low as 150 keV can be 
detected~\cite{50L}. This performance allows a detailed reconstruction  
of the solar neutrino interactions. The background induced by natural  
radioactivity and the need to establish the electron direction in  
elastic scattering events, require a threshold for the detection of  
primary electrons. As will be discussed below, this threshold is of 
the order of 5~MeV for  elastic  and absorption events, thus allowing
investigation on the higher energy part of the solar neutrino spectrum
($^8$B and hep). 
 
The performance of the ICARUS T600 detector in the analysis of solar 
neutrino events are widely discussed in a recent, dedicated ICARUS 
publication~\cite{SOLAR}. We only recall here the main results of this study,
referring to the paper for the detailed analysis, in particular regarding
the background estimate. 
 
\subsection{Event and background rates evaluation} 

ICARUS can make a fundamental contribution to our understanding of 
solar neutrino intensities and their energy spectrum, by observing 
the electron produced in the following independent processes: 
\begin{itemize}  
\item [-] elastic scattering by electrons:
$\nu _{e,\mu,\tau} + e^- \rightarrow \nu _{e,\mu,\tau} + e^-$ 
\item [-] absorption reactions on Argon nuclei:
$\nu _e + ^{40}$Ar$ \rightarrow ^{40}$K$^\ast + e^-$ 
\end{itemize}
The first step in the analysis of solar neutrino events
is the calculation of event and background rates.

\subsubsection{Intensity of neutrino events} 
 
We consider separately neutrino elastic scattering (ES) on electrons and
neutrino absorption by Argon with Fermi transition (F) to the   
4.38~MeV Isotopic Analogue State (IAS)  of $^{40}$K , and Gamow-Teller 
transitions (GT) to several $^{40}$K 
levels~\cite{Ormand:1995js}. The $^8$B solar neutrino flux used in the
calculation is taken from the BP98 standard 
solar model~\cite{Bahcall:1998wm}. 
 
The elastic scattering event rate at different values of the cutoff   
kinetic energy  of the recoil electron is computed by using the cross  
section values taken from Ref.~\cite{BOOK}.  For neutrino absorption, the  
shape of the cross section (evaluated for a transition to the IAS) is  
assumed to be the same for Fermi and Gamow-Teller transitions and the  
absolute values are computed by normalization to the theoretical values  
obtained by shell model calculations~\cite{Ormand:1995js}. 
The Fermi and Gamow-Teller contributions to the neutrino absorption  
on $^{40}$Ar can also be obtained indirectly from measurements of  
the $\beta^+$-decay of the mirror nucleus $^{40}$Ti, assuming isospin  
symmetry.  Two recent experiments have been performed. 
One of them~\cite{TRINDER}, yields cross  
section values somewhat larger, while the second~\cite{LIU} essentially  
confirms the theoretical predictions. 
The lowest cross section values have been used for the present calculations.

The resulting neutrino interaction achievable with an exposure of
1 kton $\times$ year, for ES, F and GT events, as a function
of the threshold on the primary electron kinetic energy, are summarized in  
columns 2 to 4 of table~\ref{stab1}. 
 
\begin{table}[htbp] 
\begin{center} 
\begin{tabular}{c c c c c c} 
\hline 
$T_{th}$          &\multicolumn{4}{c}{Events}             \\ 
\cline{2-6} 
(MeV)       & Elastic & Fermi & Gamow-Teller & Photons    & Neutrons \\ 
\hline 
0.0000 &  2674   &  1964  &  1902         & 1.40$\times 10^8$ &  15745   \\ 
1.0000 &  2238   &  1928  &  1902         & 3.83$\times 10^7$ &   7243   \\ 
2.0000 &  1826   &  1792  &  1868         & 2.14$\times 10^6$ &   3306   \\ 
3.0000 &  1438   &  1530  &  1832         &                   &   1481   \\ 
4.0000 &  1092   &  1151  &  1702         &                   &    677   \\ 
5.0000 &   792   &   730  &  1453         &                   &    306   \\ 
5.5000 &         &   530  &  1094         &                   &          \\ 
6.0000 &   540   &   355  &   694         &                   &    140   \\ 
6.5000 &         &   213  &   504         &                   &          \\ 
7.0000 &   347   &   111  &   338         &                   &     64   \\ 
7.5000 &         &    47  &   204         &                   &          \\ 
8.0000 &   204   &    15  &   106         &                   &     28   \\ 
8.5000 &         &     4  &    45         &                   &          \\ 
9.0000 &   106   &        &    15         &                   &          \\ 
9.5000 &         &        &     4         &                   &          \\ 
10.000 &    49   &        &               &                   &          \\ 
\hline 
\end{tabular} 
\end{center} 
\caption{Calculated solar neutrino reactions for 
an exposure of 1 kton $\times$ year, as a 
function of the primary electron kinetic energy threshold $T_{th}$.} 
\label{stab1} 
\end{table}

\subsubsection{Background estimates}

The following background sources have been considered: 
 
\begin{itemize} 
 
\item [(a)]{\em Natural radioactivity.}\hfill\break 
The decay of $^{40}$K, uranium, thorium, radon  
and daughters, present in the rock or in the atmosphere surrounding the  
detector, generate photons and can produce neutrons by spontaneous fission  
(SF) or ($\alpha, n$) reactions.\hfill\break 
 
\item [(b)]{\em Radioactive pollution in liquid Argon.}\hfill\break 
There are two Argon isotopes which are radioactive, with long life times,
$^{42}$Ar and $^{39}$Ar. While the $^{39}$Ar contamination is expected to
be minimal, $^{42}$Ar is expected to be present in atmospheric Argon
at the level of $\le 7\times 10^{-22}$ $^{42}$Ar atoms per natural 
Ar atom~\cite{AR42}. 
An experimental limit has been recently obtained of $\le 5\times 10^{-21}$ 
$^{42}$Ar atoms per natural Ar atom~\cite{AR42_exp}.
\hfill\break 
 
\item [(c)]{\em Radioactivity  of structural materials.}\hfill\break 
The materials constituting the dewar walls
have been analyzed for radioactive contamination. 
\hfill\break 
 
\item [(d)]{\em Nuclear photo-dissociation.}\hfill\break 
In addition to the natural radioactivity, high energy cosmic ray muons  
which penetrate the rock, can induce nuclear photo-dissociation, with subsequent 
neutron production. 
\end{itemize} 
 
All these various contributions have been evaluated.
Natural radioactivity  
of the rock turns out to be by far the most important background component and it 
is the only radiation source considered in our calculations.
Particular care is devoted to neutrons, which are the only radiation  
able to generate high energy electrons in  the energy  
range of the $^8$B neutrino spectrum. 
 
The calculations of the background intensities are performed by detailed 
Monte Carlo simulations. The 
purpose of these computations is to derive the background event topology and  
the frequency and energy distribution of the resulting electron tracks in  
the sensitive volume, which can fake solar neutrino events. 

The input background sources used in the Monte Carlo simulation are obtained from 
the measured photon~\cite{SOMIGLIANA} 
and neutron~\cite{NEUTRONS} spectra in the underground LNGS site. These 
were assumed to be a 
projection of the spectra of the particles emerging from the rock. The 
input neutron spectrum is the 
result of a direct measurement performed by the ICARUS collaboration 
in the Gran Sasso laboratory hall C. This measurement is in fairly good
agreement with the result of a simulation considering the uranium and thorium
specific activity of the LNGS rock~\cite{ESPOSITO}. 
Photons and neutrons are considered independently. 
 
\begin{figure}[htbp] 
\centerline{\includegraphics[width=8cm]{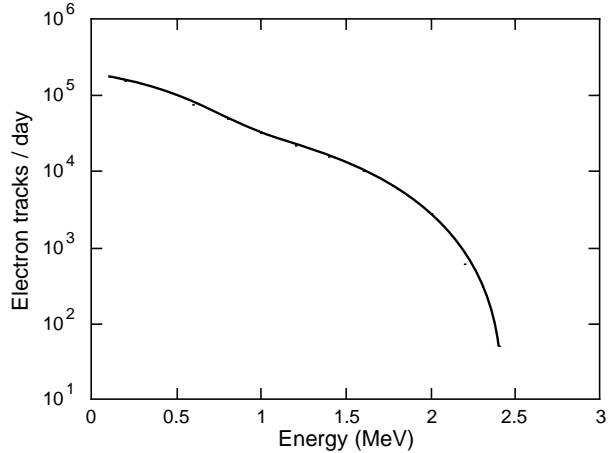}} 
\caption{Integral electron track energy spectrum generated 
from photon interaction.} 
\label{fig:sol1} 
\end{figure} 
 
\begin{itemize} 
 
\item [(a)]{\em Photons.}\hfill\break 
Photons penetrating in the detector undergo electromagnetic interactions, 
producing electrons in LAr. The energy distribution of these 
electron tracks in the detector is shown in figure~\ref{fig:sol1}.   
No electron tracks are expected with an energy larger than 2.4~MeV. 
Each event consists of a main track, possibly accompanied by electrons  
produced by the interaction of bremsstrahlung photons. The distribution of  
these events, according to the energy of the most energetic electron 
(primary track),  
is displayed in the  fifth column of table~\ref{stab1}.\hfill\break 

\item[(b)]{\em Neutrons.}\hfill\break 
Neutrons entering the detector may undergo neutron-capture followed by
gamma ray emission. Gamma interactions produce electron tracks. 
The computation is  
performed in two steps. First, the calculation of the 
neutron capture intensity, 
occurring mostly in LAr, is performed by the MCNP code. 
The simulated ICARUS detector consists
of an external neutron shielding layer, a  
thermal insulating material cavity, a liquid Argon dead region (0.35~m  
thick), and the inner sensitive volume. 
Second, the  abundance of the events generated by the gamma rays 
and classified according to their nature 
(electron tracks multiplicity, energy, etc.) was calculated using a 
GEANT based simulation. The event topology after
capture consists of a number of electron tracks  
produced by de-excitation or by  bremsstrahlung photons. From this
calculation 
about 15745 neutrino capture events are expected for an exposure 
of 1 kton $\times$ year. 
The distribution of these events, according to 
the energy of the most energetic 
electron  track, is displayed  in the last column of table~\ref{stab1}. 

The topologies of the neutron capture events, in which at least one  
electron has kinetic energy larger then 5~MeV (obtained with the GEANT  
program) are displayed in table~\ref{stab2}~d).  
Here, the fraction of events is shown  
as a function of the Compton electron multiplicity and its 
associated energy (total energy of secondary electrons). 
This table, together with the values displayed in the last column of  
table~\ref{stab1}, will allow the computation of the background 
contamination in each class of events.
\end{itemize} 
 
\subsection{Solar neutrinos detection} 
 
The choice of a primary electron kinetic energy threshold of 5~MeV, to 
select the solar neutrino event sample, is justified by
statistical considerations on neutrino signal versus background event rates
(reported in table~\ref{stab1}).
With this selection there is no photon background contribution, while a
total sample of 2975 solar neutrino events is expected (for 
1 kton $\times$ year 
exposure), with a contamination of 306 events from neutron background.

In order to define off-line event versus background 
selection criteria, a full GEANT
Monte Carlo simulation, which performs the transport of gamma rays and 
electrons inside the liquid Argon, has been carried out.  Every
electron track is then digitized by the following procedure: 
 
\begin{itemize} 
\item[(a)] The deposited energy is converted in charge. 
\item[(b)] The charge is drifted towards the anode of the chamber, with an 
infinite electron life time in LAr. 
\item[(c)] Digitized electronic signals are generated on three wire planes  
(forming the anode), placed at $60^\circ$ from one another, with 3~mm wire pitch. 
\item[(d)] Gaussian distributed electronic noise is added with zero mean  
value and a RMS 1000 electrons. The resulting electron 
threshold, which is strongly correlated with electronic noise and with the 
sense wire pitch, is 150~keV. 
\end{itemize} 
The digitized signals are picked up from the noise by means of an  
integral-differential algorithm and
the final parameters (position and energy after digitization) are  
obtained by a fitting procedure of the signals, equivalent to the
procedures used for the analysis of real events.
 
\begin{figure}[htbp] 
\centerline{\includegraphics[width=13cm]{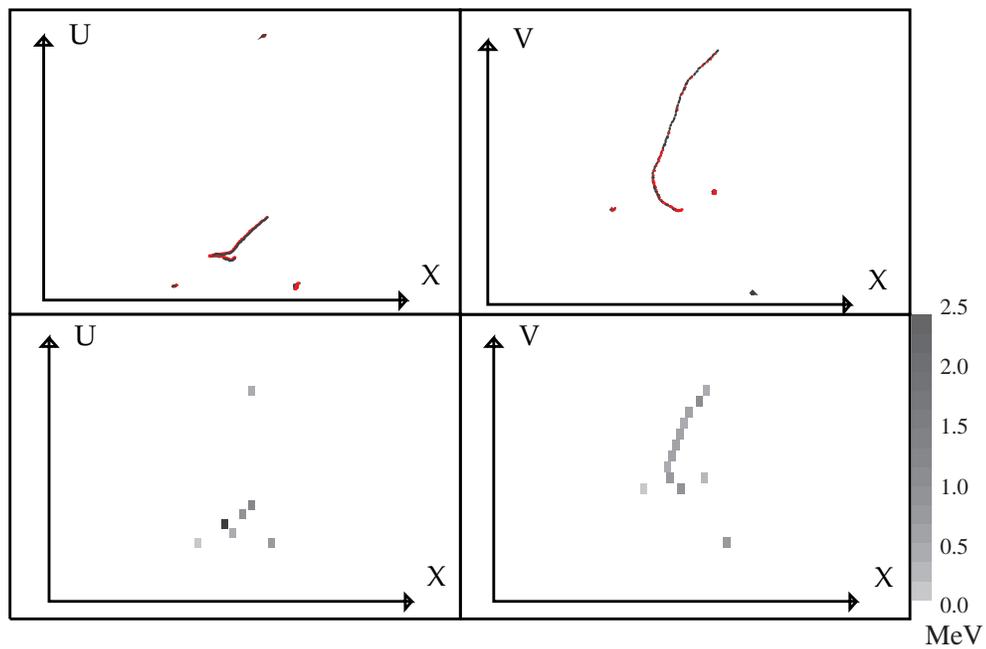}} 
\caption{Top figure: an absorption event as simulated by the GEANT Monte Carlo 
program is shown in two wire planes (U and V coordinates) put at an angle of  
$60^\circ$, the X axis is the drift coordinate. The projected track length 
is about 3~cm, the main electron energy is 7~MeV, the associated  
energy is 2~MeV  and the Compton electron multiplicity is 3. At the  
bottom the same event is shown after digitization and signal extraction. 
The grey scale of each  
pixel is proportional to the deposited charge. The resolution in the  
horizontal axis (drift direction) is  
0.4~mm, and in the vertical axis is 3~mm (wire pitch).} 
\label{fig:sol3} 
\end{figure}

As an example, a Monte Carlo absorption event is shown in 
figure~\ref{fig:sol3}.  It is  
characterized by the track of the primary electron generated in the  
interaction, surrounded by a number of secondary tracks produced by  
photons following the $^{40}$K$^\ast$ de-excitation. 

The Monte Carlo simulation of ES and absorption events shows that the 
probability of finding  secondary  
electron tracks vanishes  50~cm  away from the interaction point. 
Electrons undergo strong multiple scattering and bremsstrahlung.   
Deviation from straight line tracks decreases with increasing energy 
(due to the 1/p dependence showed by multiple scattering):   
track directions for electron energies larger than 5~MeV are efficiently
reconstructed.
 
The correlation between Compton electron multiplicity and its
associated energy  
will be used to define the off-line event selection criteria and to evaluate 
the selection efficiencies $\epsilon_{ES}$, $\epsilon_{GT}$  
and $\epsilon_{F}$ for scattering and absorption channels, and the
background rejection power.
 
\subsubsection{Elastic scattering} 
 
\begin{figure}[htbp] 
\centerline{\includegraphics[width=9cm]{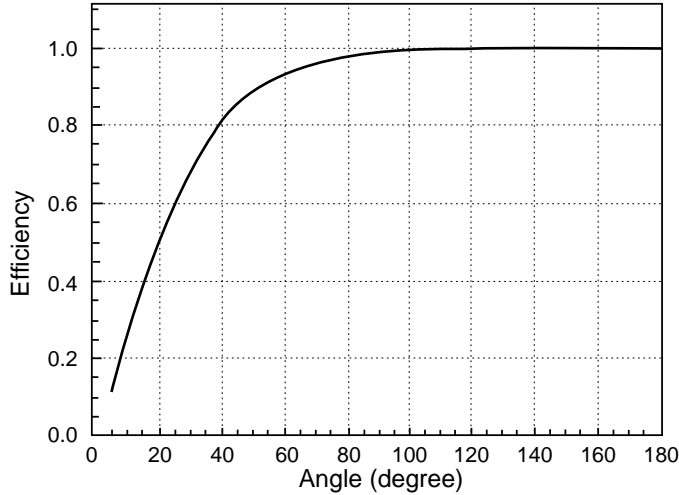}} 
\caption{Fraction of elastic events ($E > 5$~MeV) 
as a function of the cone angle  
within which the reconstructed electron direction in space  is contained; 
the cone axis is defined by the parent  neutrino direction. 
Three wire planes put at $60^\circ$ angle 
and a 3~mm wire pitch are used.} 
\label{fig:sol4} 
\end{figure}

Electrons produced via an ES have an angular  
distribution, which is strongly peaked forward with respect to the 
initial solar neutrino direction, as shown in figure~\ref{fig:sol4}. 
This feature provides an efficient mean to distinguish neutrino 
elastic events from
background, assuming that the electron vertex is correctly reconstructed from 
the energy deposited along the track. In our computations 
the first hit wire can be distinguished from the end point of one electron  
with an efficiency larger than 80\%. 
Moreover, electrons 
from ES reactions are essentially isolated. The fraction of the ES events as a 
function of the multiplicity  
and energy of the secondary tracks is shown in table~\ref{stab2}~a). 

\begin{table}[htbp] 
\begin{center} 
\centerline{a) Elastic scattering events}
\begin{tabular}{c c c c c} 
\hline 
Associated &\multicolumn{4}{c}{Compton electron multiplicity}\\ 
\cline{2-5} 
Energy     & 0 & 1 & 2 & 3 \\ 
\hline 
$E<1$~MeV    & 0.880 & 0.073 & 0.008 & 0 \\ 
$E\ge 1$~MeV & 0     & 0.015 & 0.015 & 0.009 \\ 
\hline 
\end{tabular} 
\end{center} 

\begin{center} 
\centerline{b) Gamow-Teller events}
\begin{tabular}{c c c c c} 
\hline 
Associated &\multicolumn{4}{c}{Compton electron multiplicity}\\ 
\cline{2-5} 
Energy     & 0 & 1 & 2 & $\ge 3$ \\ 
\hline 
$E<1$~MeV & 0.083 & 0.168 & 0.049 & 0 \\ 
$E\ge 1$~MeV   & 0     & 0.075 & 0.297 & 0.328 \\ 
\hline 
\end{tabular} 
\end{center} 

\begin{center} 
\centerline{c) Fermi events}
\begin{tabular}{c c c c c} 
\hline 
Associated &\multicolumn{4}{c}{Compton electron multiplicity}\\ 
\cline{2-5} 
Energy     & 0 & 1 & 2 & $\ge 3$ \\ 
\hline 
$E<1$~MeV    & 0.032 & 0.039 & 0.018 & 0 \\ 
$E\ge 1$~MeV & 0     & 0.081 & 0.221 & 0.519 \\ 
\hline  
\end{tabular} 
\end{center} 

\begin{center} 
\centerline{d) Neutron capture events}
\begin{tabular}{c c c c c} 
\hline 
Associated &\multicolumn{4}{c}{Compton electron multiplicity}\\ 
\cline{2-5} 
Energy     & 0 & 1 & 2 & $>2$ \\ 
\hline 
$E<1$~MeV    & 0.46 & 0.26 & 0.10 & 0   \\ 
$E\ge 1$~MeV & 0    & 0.05 & 0.07 & 0.06 \\ 
\hline 
\end{tabular} 
\end{center} 
\caption{Fraction of events with at least one 
electron with kinetic energy larger than 5~MeV, as a function of the 
Compton electron multiplicity and its associated energy. Data obtained 
after digitization are used.} 
\label{stab2} 
\end{table} 

Therefore, we assume the following off-line selection criteria for ES events: 
\begin{enumerate} 
\item primary electron energy larger than 5~MeV; 
\item cone aperture around the direction of the emitted electron of $25^\circ$:
efficiency $\epsilon_1 = 0.65$; 
\item Compton electron multiplicity M=0: efficiency 
$\epsilon_2 = 0.88$, see table~\ref{stab2}~a). 
\end{enumerate} 
The total detection efficiency is $\epsilon_{ES}=0.572$, corresponding to an
ES expected rate of 453 events, for 1 kton $\times$ year exposure 
(see table~\ref{stab1}). 

The contamination of neutron capture events is 306 events (see  
table~\ref{stab1}). Applying the ES selection criteria on the neutron sample
(angular cut efficiency
$\epsilon_1 = 0.047$, multiplicity cut $\epsilon_2 = 0.46$, using 
table~\ref{stab2}~d)), we obtain 7 residual background events. If we are not
able to determine the electron direction of flight, the angular cut efficiency
is half and we remain with 14 background events.

\subsubsection{Absorption events} 

The angular distributions of electrons produced by absorption events,
to a first approximation,
can be considered isotropic, therefore angular cuts are not effective.
Efficient discrimination criteria are based on associated energy and 
multiplicity cuts. 
 
\begin{itemize} 
\item[(a)]{\em Allowed Gamow-Teller transitions.}\hfill\break 
In table~\ref{stab2}~b) the  
correlation between the associated multiplicity and secondary electron 
total energy for GT events is shown. 
We assume the following off-line selection criteria for the GT sample: 
\begin{enumerate} 
\item primary electron kinetic energy larger than 5~MeV; 
\item associated energy $E\ge 1$~MeV and multiplicity $M\ge 1$: 
$\epsilon_{GT} = 0.70$, see table~\ref{stab2}~b). 
\end{enumerate} 
The final expectation after cuts is 1017 events, for 1 kton $\times$ year exposure.
\hfill\break 
 
\item[(b)]{\em Super allowed Fermi transition.}\hfill\break 
In table~\ref{stab2}~c) the 
correlation between the associated multiplicity and energy is shown.  

We assume the following off-line selection criteria for F events:
\begin{enumerate} 
\item primary electron kinetic energy larger than 5~MeV, 
\item associated energy $E\ge 1$~MeV and multiplicity $M \ge 1$:
$\epsilon_{F}=0.82$, see table~\ref{stab2}~c). 
\end{enumerate} 
The final expectation after cuts is 599 events, for 1 kton $\times$ year exposure. 
\end{itemize} 
 
The total absorption rate (GT+F) is about 1616 events, for 
1 kton $\times$ year exposure,
with a contamination from neutron captures of 55 events, applying the absorption
cuts to the neutron sample (see table~\ref{stab1} and \ref{stab2}~d)). 

\vspace{0.5 cm}
Besides the neutron contamination, 8.3\% of the GT  and 3.2\% 
of  the F type events can fake an 
elastic scattering event, as can be seen in tables~\ref{stab2}~b) and 
\ref{stab2}~c) (multiplicity M=0). Taking into account the angular cut, this 
reduces the contamination to 11 events, for 1 kton $\times$ year exposure, in the ES 
sample. The contamination of the absorption sample from ES is  
3.9\%, i.e. 17 events, for 1 kton $\times$ year exposure, as can be derived from 
tables~\ref{stab1} and \ref{stab2}~a) (multiplicity $M \ge 1$ and associated 
energy $\ge 1$~MeV). 
 
\subsection{Sensitivity to oscillations} 
 
We summarize the results in table~\ref{stab6}, where the estimated rates of  
events per year are shown, together with the background rates. 
From this table we conclude that a clean measurement of $^8$B solar  
neutrinos can be performed in a reasonable data-taking time.  
Direct proof of the oscillation mechanism will be possible for 
a vast fraction of the presently allowed parameter region through the 
comparison of the elastic and absorption event rates (table~\ref{stab7}). 
An accurate measurement of the $^8$B spectrum will also be performed by 
means of the absorption event sample. It is  
important to bear in mind that this is possible because of the low  
intensity of background signals.  Noise is mainly related to the neutron  
flux level in the LNGS laboratories, but the radioactive contamination 
of the materials employed in the detector construction must be accurately considered. 
At present in one of the half-modules the electric-insulation 
material is not adequate from the point of view of the radiochemical 
purity. We estimate that this material will induce an increase of 
the background by a factor of 10. However, we plan to substitute it before 
the installation in the underground lab. 
 
\begin{table}[htbp] 
\begin{center} 
\begin{tabular}{l c} 
\hline 
                                &  Events/year \\ 
\hline 
Elastic channel ($E \ge 5$~MeV) & 453 \\ 
Background                      &  14 \\ 
Absorption event contamination  &  11 \\ 
\hline 
Absorption channels             & 1616 \\ 
Background                      &   55 \\ 
Elastic event contamination     &   17 \\ 
\hline 
\end{tabular} 
\end{center} 
\caption{Number of events expected with an exposure of 1 kton $\times$ year, 
compared with the computed background (no oscillation).} 
\label{stab6} 
\end{table} 

\begin{table}[htbp] 
\begin{center} 
\begin{tabular}{c c c c} 
\hline 
Solution region  & $R$   & Exclusion level & Minimal exposure  \\ 
                 &       &                 & (kton $\times$ year) \\ 
\hline 
\hline 
Active MSW - SMA          & $1.0\div 1.1$ & Nearly complete & 2 \\ 
\hline 
Active MSW  & $1.1\div 1.3$ & Only the largest & $>0.5$\\ 
Extended SMA &               & mixing side      & \\ 
\hline 
Active MSW - LOW & $1.1\div 1.3$ & Complete & $>0.5$ \\ 
\hline 
Active MSW - LMA & $1.3\div 1.9$ & Complete & $0.5$ \\ 
\hline 
Active MSW  & $1.2\div 2.3$ & Complete & $0.5$ \\ 
Extended LMA & & & \\ 
\hline 
Active JustSo & $0.8\div 1.0$ & Partial & $>0.5$ \\ 
\hline 
Active JustSo & $1.0\div 2.0$ & Partial & $>0.5$ \\ 
\hline 
Sterile MSW - SMA &$0.8\div 0.9$ & Complete & 0.5 \\ 
\hline 
Sterile MSW  &$0.6\div 0.8$ & All the higher & 0.5 \\ 
Extended SMA  &              & mixing angle side & \\ 
\hline  
Sterile MSW  &$0.6\div 1.0$ & Complete & $>0.5$ \\ 
Extended SMA & & & \\ 
\hline 
\end{tabular} 
\end{center} 
\caption{The $R$-ratio range, the level of exclusion and the minimal 
exposure (units of 1 kton $\times$ year).} 
\label{stab7} 
\end{table} 
 
In Super-K about 13~solar neutrino events/day (elastic scattering),
with an energy threshold of 5 MeV, are observed. 
In the ICARUS T600 detector, with a neutrino oscillation hypothesis,  
1.4~solar neutrino events/day (elastic scattering + absorption) with 
$E > 5$~MeV are expected.
Therefore, the statistical accuracy attainable with the T600 experiment
is by far worse compared to Super-K.  However, systematic uncertainty 
is expected to be lower, due to the higher event selection efficiency 
and energy resolution. More important, with ICARUS one can exploit 
the two available solar neutrino interaction processes. This is relevant
to enhance the sensitivity to oscillation.

Recently the SNO experiment started operation (November 1999). The active 
mass is about 1 kton of heavy water ($D_2 O$) and the energy threshold is
5 MeV. The main solar neutrino detectable reactions are CC neutrino absorption
by deuterium and NC neutrino dissociation of deuterium. In case of no oscillation,
the two reactions follow the CC:NC=2.05:1.00 ratio. The expected number of
CC reactions is about $9 \times 10^3$ events/year, without taking into account 
efficiency cuts, depending on detector performance and varying with possible 
oscillation parameters~\cite{Bahcall_SNO}.  At present (phase 1) only CC reactions 
can be detected, NC reaction detection will be possible in phase 2, when salt 
will be diluted in $D_2 O$. Just as for radiochemical solar neutrino 
experiments, there is no energy discrimination for the NC reaction. Oscillation
analysis is performed by the NC/CC ratio, i.e. with an approach  similar to
the one foreseen with ICARUS.
 
There is no doubt that the quality of the data and the extra measurements 
done with the T600 detector (i.e., with a different experimental technique) 
will be a major contribution to solar neutrinos understanding. 
 
\subsection{Oscillation parameters allowed regions} 
 
A possible way to combine the ICARUS measurements from the two  
independent detection channels, elastic scattering and absorption  
events (Gamow-Teller and Fermi),  is to compute the following ratio: 

\begin{equation} 
R=\frac{\frac{N^{ES}}{N^{ES}_0}} 
{\frac{1}{2}\left( \frac{N^{GT}}{N^{GT}_0}+\frac{N^{F}}{N^{F}_0}\right) } 
\end{equation} 
where $N^{ES}$, $N^{GT}$, $N^{F}$ are the measured event rates 
(elastic, Gamow-Teller and pure Fermi respectively), 
and $N_0^{el}$, $N_0^{GT}$, $N_0^F$ 
are the predicted event rates in the case of standard neutrino without  
oscillations. 
 
The proposed ratio is an indicator  with the following advantages: 
\begin{itemize} 
\item  it is independent of the $^8$B total neutrino flux, predicted by  
different solar models, and of any possible pure astrophysical suppression  
factor; 
\item it does not depend on experimental threshold energies or on the  
adopted cross-sections for the different channels. 
\end{itemize} 
 
The quantities introduced above are defined as follows.  
 
\begin{equation} 
N^{ES}= \Phi^{SM}_{^8{\rm B}}\int_{E_{\nu,{\rm min}}}^{+\infty} dE_\nu 
\, S(E_\nu)\, \left[ \sigma^{ES}_{\nu_e}(E_\nu)\, P(E_\nu) + 
\sigma^{ES}_{\nu_{\mu(\tau)}}(E_\nu)\, \left( 1-P(E_\nu) \right) \right]  
\end{equation} 
here $E_\nu$ is the neutrino energy, $S(E_\nu)$ is the standard $^8$B  
neutrino spectrum, $\sigma_{\nu_e}^{ES}(E_\nu)$ 
is the elastic scattering cross-section for electron-neutrinos while   
$\sigma_{\nu_{\mu(\tau)}}^{el}(E_\nu)$ 
is the corresponding cross-section for mu-neutrinos 
or tau-neutrinos and 
$P(E_\nu)$ is the survival probability for 
$\nu_e \to \nu_{\mu(\tau)}$  or  $\nu_e \to \nu_S$ transitions.  
In the second case the   contribution has to be omitted.  These  
probabilities are a function of neutrino parameters  
$\Delta m^2$ and $\sin^2 2\theta$. The lower limit in the integral is 
\begin{equation} 
E_{\nu,{\rm min}} = \frac{1}{2} \left[ T_{th}+ 
\sqrt{T^2_{th} + 2T_{th}\mbox{m}_e} \right] 
\end{equation} 
where $T_{th}$ is the electron threshold kinetic energy and  
$m_e$ is the electron mass. 
 
For Gamow-Teller and Fermi transitions the corresponding event rates  
are defined as: 
\begin{equation} 
N^{GT}= \Phi^{SM}_{^8{\rm B}}\int_{E_{\nu,{\rm min}}^{GT}}^{+\infty} dE_\nu 
\, S(E_\nu)\, \sigma^{GT}(E_\nu)\, P(E_\nu) 
\end{equation} 
where 
\begin{center} 
$E_{\nu,{\rm min}}^{GT} = T_{th}$ + 1.50 MeV + 2.29 MeV 
\end{center} 
and 
\begin{equation} 
N^{F} = \Phi^{SM}_{^8{\rm B}}\int_{E_{\nu,{\rm min}}^{F}}^{+\infty} dE_\nu 
\, S(E_\nu)\, \sigma^{F}(E_\nu)\, P(E_\nu) 
\end{equation} 
where 
\begin{center} 
$E_{\nu,{\rm min}}^{F} = T_{th}$ + 1.50 MeV + 4.38 MeV. 
\end{center} 
 
The corresponding neutrino event rates without oscillation 
($N_0^{ES}$, $N_0^{GT}$, $N_0^{F})$ may  
be obtained from the previous formulae, putting  $P(E_\nu \equiv 1)$. 

\begin{figure}[htbp] 
\vspace{1cm}
\centerline{\includegraphics[width=13cm]{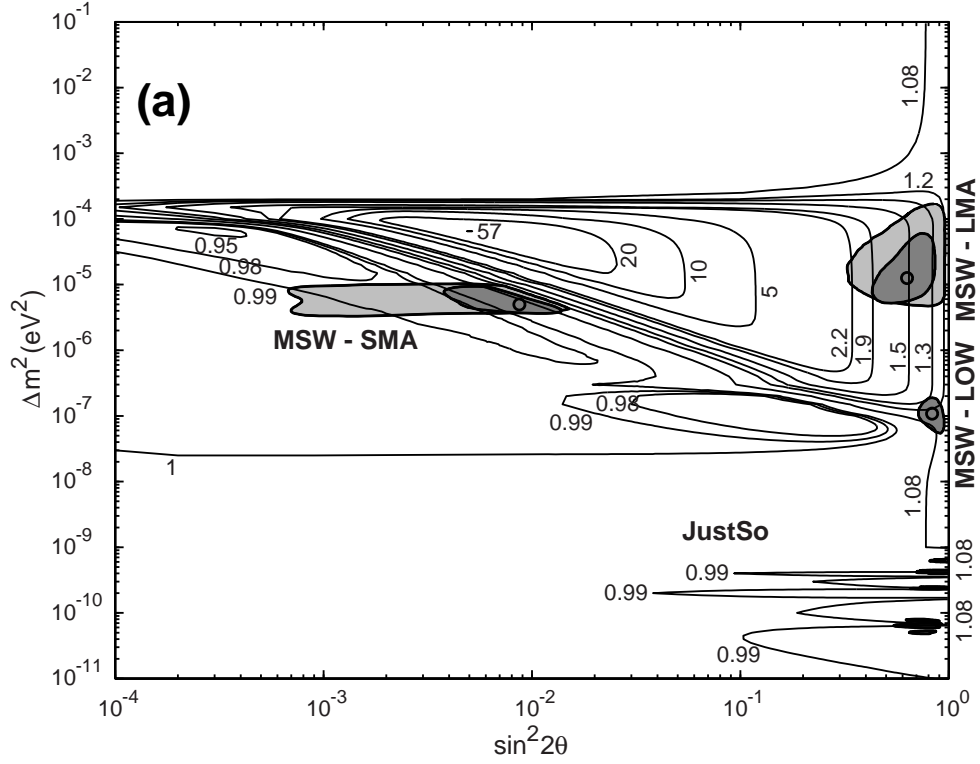}} 
\caption{Iso-$R$ curves for active neutrinos.} 
\label{fig:sol7} 
\end{figure} 
 
The iso-$R$ curves for active neutrinos obtained from Monte Carlo 
simulation, taking into  
account neutrino oscillation, are shown in figure~\ref{fig:sol7}. 
The corresponding study for sterile neutrinos is reported in~\cite{SOLAR}. 
 
Shaded regions represent the allowed regions resulting  
from five solar neutrino experiments (Homestake-chlorine, Kamiokande, 
GALLEX, SAGE, and Super-Kamiokande) for 90\% and 95\% confidence level. 
The solutions of the SNP (which the ICARUS   
experiment will be able to probe) together with the  
$R$-ratio   
range, the level of exclusion, and the minimal exposure are reported 
in table~\ref{stab7}. 
 
There are combinations of neutrino parameters for which $R=1$,  
despite of the fact that the MSW effect takes place 
in the sun. Such regions can not 
be excluded by this method and represent its theoretical limit. 
The experimental limit is given by the ICARUS ability to detect a  
small deviation from $R=1$, which is related to the statistical 
error and  all the experimental systematic error sources. 
 
\begin{table}[htbp] 
\begin{center} 
\begin{tabular}{c c c c} 
\hline 
Exposure (kton $\times$ year) & $\Delta R/ R$ \% & $R_{min}$ & $R_{max}$ \\ 
\hline  
1 &  7.5 & 0.92 & 1.08 \\ 
2 &  5.3 & 0.95 & 1.05 \\ 
4 &  3.8 & 0.96 & 1.04 \\ 
\hline 
\end{tabular} 
\end{center} 
\caption{One-sigma relative uncertainty for $R$, as a function of the 
exposure, and limits of the one-sigma exclusion region.} 
\label{stab8} 
\end{table} 
 
Taking into account only the statistical error  and the rates  
reported above, we can estimate the one-sigma relative uncertainty 
for $R$, as a function of the exposure  
time. The results are given in table~\ref{stab8}. 
 
From the results above it is clear that it will be possible to test
some of the currently allowed solutions with an exposure limited
to 0.5-2 kton $\times$ year.


\section{Nucleon decay searches}
\label{sec:nucdecay}

\def  \pepi      {$p \rightarrow e^+  \pi^0$}
\def  \penopi    {$p \rightarrow e^+   \; ( \pi^0 )$}
\def  \pkanu     {$p \rightarrow \bar{\nu} K^+ $}
\def  \pmupik    {$p \rightarrow \mu^- \; \pi^+ \; K^+$}
\def  \pepipi    {$p \rightarrow e^+   \;   \pi^+ \;   \pi^-   $}
\def  \pepinopi  {$p \rightarrow e^+   \;   \pi^+ \; ( \pi^- ) $}
\def  \penopipi  {$p \rightarrow e^+   \; ( \pi^+ \;   \pi^- ) $}
\def  \ppianu    {$p \rightarrow \pi^+ \; \bar{\nu} $}
\def  \pmupi     {$p \rightarrow \mu^+ \;   \pi^0   $}
\def  \pmunopi   {$p \rightarrow \mu^+ \; ( \pi^0 ) $}
\def  \nek       {$n \rightarrow e^-   \; K^+$}
\def  \nepi      {$n \rightarrow e^+   \;   \pi^-$}
\def  \nenopi    {$n \rightarrow e^+   \; ( \pi^- )$}
\def  \nmupi     {$n \rightarrow \mu^- \;   \pi^+$}
\def  \nmunopi   {$n \rightarrow \mu^- \; ( \pi^+ )$}
\def  \npianu    {$n \rightarrow \pi^0 \; \bar{\nu}$}

The question of baryonic matter stability is of paramount importance, 
since proton decay offers an unique way to have an insight of what
happens beyond what currently appears to be the desert after the
standard model. The theoretical ideas and models relevant to proton
decay require some new, very high, intermediate mass scale. Such a
mass scale will never be reached with today's acceleration
techniques. Non-accelerator experiments are the only way to
explore experimentally the phenomenology at such high energies.

The Super-Kamiokande collaboration has extensively probe the classical
decay channels (e.g., \pepi) and the dominant decay
mode, \pkanu\, according to SUSY Grand Unified Theories 
(see e.g. Ref.~\cite{sk-pdk2000}).
Plans exist to operate a megaton water \v{C}erenkov
detector~\cite{Jung:1999jq} in view of improving current sensitivities for 
 \pepi\ and \pkanu\ modes by at least one order of magnitude.
However, to unmistakably show the existence of a signal, 
these experiments have to rely on statistical background subtraction.

A clear advantage and certainly the main strength of the ICARUS
technique is, that discovery will be
possible at the level of a single event, 
thanks to its superb imaging and energy 
resolution capabilities. In addition, a full understanding of the 
mechanism responsible for proton decay, 
requires a precise measurement of all possible branching ratios. 
Since ICARUS provides a much more powerful background 
rejection, it can perform a large variety of exclusive decay 
modes measurements. Inclusive searches are obviously also possible.

Hence, a liquid Argon detector is an ideal device, in particular, 
for those channels that 
are not accessible to \v{C}erenkov 
detectors due to the complicated event topology, or because the emitted 
particles are below the \v{C}erenkov threshold (e.g. $K^{\pm}$). In
particular, the operation of a T600 module, at the
Gran Sasso laboratory, will be of the utmost importance to verify
both, the predicted background levels and the anticipated detector efficiencies.

\begin{figure}[htbp]
\centering
\epsfig{file=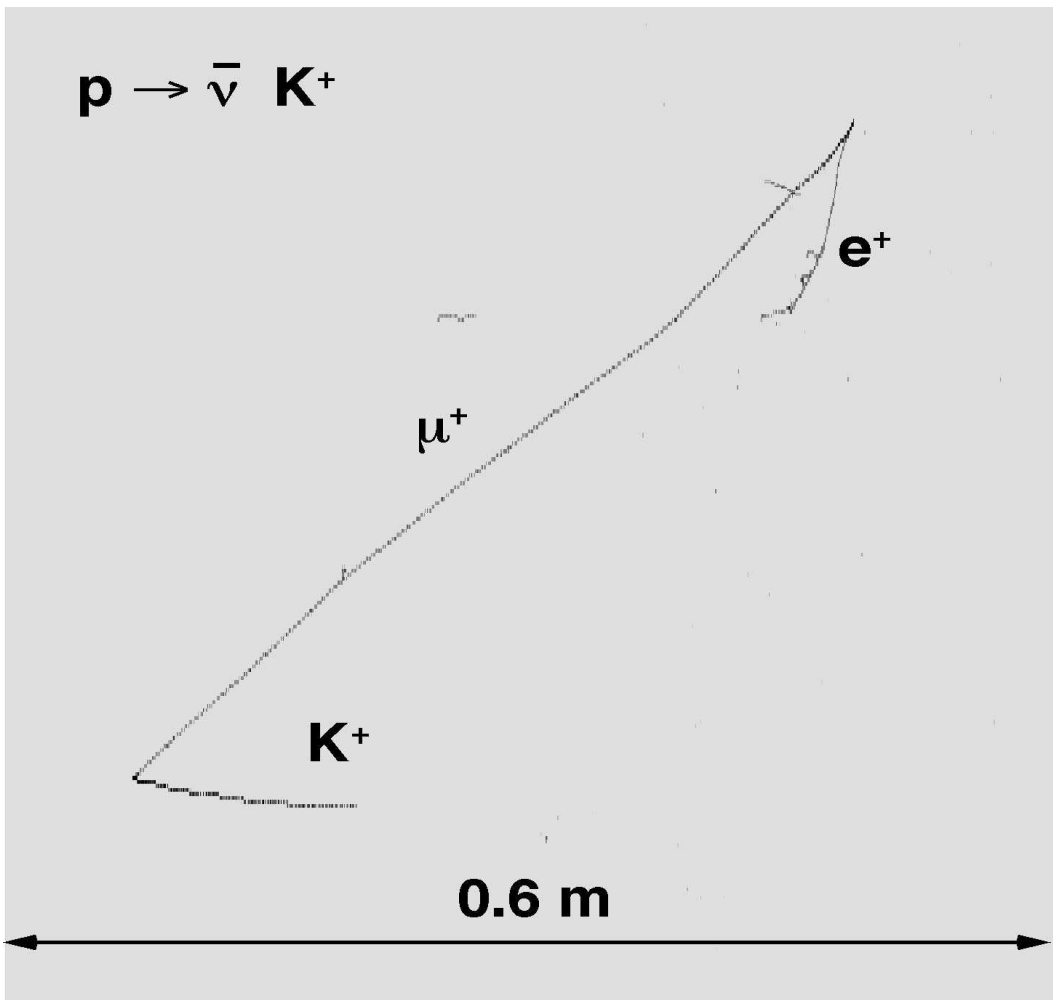,width=14cm}
\caption{Simulated proton decay in the preferred channel in
Supersymmetric models
$p \rightarrow \bar\nu K^{+}$ as could be observed in \ICARUS.}
\label{fig:nuk}
\end{figure}

We have performed a detailed full event simulation based on the FLUKA
package~\cite{flukakek, flukalisb} and the realistic events obtained contain very 
long tracks with redundant information, allowing particle identification and 
measurement of their energies with great precision. See, for instance, the 
spectacular example of the SUSY-preferred decay mode of the proton 
$p \rightarrow \bar\nu K^{+}$ displayed in Figure~\ref{fig:nuk}. 
We can observe the increase in ionization 
deposition by the $K^{+}$ as it comes to rest. There is no ambiguity in the direction 
of the particle along its trajectory. Particle
identification benefits greatly from the ability to measure the 
ionization loss ($dE/dx$). In particular, using $dE/dx$ versus range only, an 
excellent separation is obtained between pions and
kaons.

We report here the study for 
two specific channels: ${\rm p \rightarrow e^{+} \pi^{0}}$ and 
${\rm p \rightarrow \bar\nu K^+}$~\cite{stony}.
The estimation of backgrounds, due to atmospheric neutrino interactions, has
been carried out using a statistical sample which is a thousand times larger than
the one expected for a 1 kton $\times$ year exposure.

\subsection{${\rm p \rightarrow e^{+} \pi^{0}}$ decay mode}

This is one of the mode favored by minimal SU(5) theories.
The list of cuts used for the exclusive scenario search is presented in
table~\ref{tab:1}, normalized to an exposure of 1 kton $\times$
year. In the presence of nuclear matter, pion induced reactions are 
complex, mainly because of two and three nucleon absorption
processes. Pion-nucleon interactions can proceed through the
non-resonant and the p-wave channels with the formation of a $\Delta$
resonance. The $\Delta$ can either decay (resulting in elastic
scattering or charge exchange), or interact with surrounding nucleons,
resulting in pion absorption. In s-wave absorption, the relative
probability of absorption on a np pair or on a nn or pp pair, is
assumed to be the same as in p-wave absorption. 
Our nuclear interaction generator~\cite{flukakek, flukalisb} foresees that 
$\sim$45\% of the times the $\pi^0$ is absorbed and therefore not visible. 
The idea behind the kinematic cuts is to have a balanced event, with
all particles identified, and with a total invariant mass compatible
with that of a proton.

{\small
\vspace{0.1cm}
\begin{table}[htbp]
\begin{center}
\begin{tabular}{|c|c|c|c|c|c|c|}\cline{2-7}
\multicolumn{1}{c|}{\bf{ Exclusive Channel Cuts}} & \bf{$p \rightarrow e^+     \pi^0               $} &
\bf{$\nu_e$ CC}& \bf{$\bar{\nu}_e$ CC} & \bf{$\nu_{\mu}$ CC}&
\bf{$\bar{\nu}_{\mu}$ CC} & \bf{$\nu$ NC} \\
\hline
One $\pi^0$                              &  54\% &  6.6 &  2.1 & 15 &  5.8 &  11.1 \\ \hline
One electron                             &  54\% &  6.6 &  2.1 & 0.02 &     0 &     0 \\ \hline
$E^{kinetic}_{proton} <$ 100 MeV                   &  52\% &  2.7 &  1.4 & 0.004 &     0 &     0 \\ \hline
$0.93< E_{total} < 0.97$ GeV        &  38\% &  0.03 & 0.01 &     0 &     0 &     0 \\ \hline
$P_{total}< 0.46$ GeV              &  37\% &  0 &     0 &     0 &     0 &     0 \\ \hline
\hline
\end{tabular}
\vspace{0.1cm}
\caption{\small Cuts for the $p \rightarrow e^+     \pi^0               $
 channel. Survival fraction of 
 signal (first column) and background events through event 
 selections applied in succession (normalized to an exposure of 1
kton $\times$ year).}
\label{tab:1}
\end{center}
\end{table}
}

  After asking for one electron, only the $\nu_e$ CC and $\bar{\nu}_e$ CC 
backgrounds survive. The key cut is the total visible energy.
A cut on the total visible energy is actually more
efficient rejecting background than a cut on the invariant mass. 
No background events are expected for an overall predicted efficiency of about 37\%.

\subsection{${\rm p \rightarrow \bar\nu K^+}$  decay mode}

  This is one of the favored SUSY decays, and is special because of the
presence of a strange meson in the final state. The conservation of
strangeness leads to very different interactions of the $K^+$, $K^0$
and $K^-$, $\bar K^0$ with nucleons at low energies. $K^-$ have a
large cross section for hyperon production, with the $\Sigma \pi$ and
$\Lambda\pi$ channels always open. They feel, like pions, a strong
nuclear potential, and therefore are strongly absorbed by nuclei. On
the other hand, $K^+$ interact relatively weakly, therefore the
fraction of positive kaons absorbed by nuclei is at the level of a few
per cent.

ICARUS profits
from its very good particle identification capabilities to tag 
the kaon and its decay products. Applying the cuts listed in 
table~\ref{tab:3}, a very good efficiency
of 97\% is reached for a negligible background. One topological cut
(ask for the presence of only one kaon in the event) and one
kinematic cut on the total energy are sufficient to obtain the quoted result.

{\small 
\vspace{0.3cm}
\begin{table}[htbp]
\begin{center}
\begin{tabular}{|c|c|c|c|c|c|c|c|}\cline{2-8}
\multicolumn{1}{c|}{\bf{Cuts}} & \bf{$p \rightarrow \bar{\nu}  K^+             $} &
\bf{$\nu_e$ CC}& \bf{$\bar{\nu}_e$ CC} & \bf{$\nu_{\mu}$ CC}&
\bf{$\bar{\nu}_{\mu}$ CC} & \bf{$\nu$ NC} & \bf{$\bar{\nu}$ NC} \\
\hline
One Kaon                                 &  97\% &   0.310 &    0.059 &   0.921 &   0.214 &   0.370 &   0.104 \\ \hline
No $\pi^0$                               &  97\% &   0.161 &    0.030 &   0.462 &   0.107 &   0.197 &    0.051 \\ \hline
No electrons                             &  97\% &     0 &     0 &   0.455 &   0.107 &   0.197 &    0.051 \\ \hline
No muons                                 &  97\% &     0 &     0 &     0 &     0 &   0.197 &    0.051 \\ \hline
No charged pions                         &  97\% &     0 &     0 &     0 &     0 &   0.109 &    0.022 \\ \hline
$E_{total} <$ 0.8 GeV                 &  97\% &     0 &     0 &     0 &     0 &     0 &     0 \\ \hline
\hline
\end{tabular}
\vspace{0.1cm}
\caption{\small Cuts for the $p \rightarrow \bar{\nu}  K^+              $
 channel. Survival fraction of 
 signal (first column) and background events through event 
 selections applied in succession (normalized to an exposure of 1 kton
$\times$ year).}
\label{tab:3}
\end{center}
\end{table}
}

\subsection{Sensitivity to nucleon decay}

To calculate partial lifetime lower limits, $(\tau / B)$, we use
the following formulae:

\[(\tau / B)_p > \frac{2.69}{S} \times Expo \times \epsilon \times
10^{32} \; \; yrs \; \; \; \; {\rm (proton \ decay)} \]
\[(\tau / B)_n > \frac{3.29}{S} \times Expo \times \epsilon \times
10^{32} \; \; yrs \; \; \; \; {\rm (neutron \ decay)} \] \\

Here, $Expo$ is the full detector exposure in kilotons per year,
$\epsilon$ is the selection efficiency, and $S$ is the constrained
90\% CL upper limit on the number of observed signal events.

$S$ is found by solving the equation:

\[ \frac{\sum_{n=0}^{n_0} P(n,b+S)}{\sum_{n=0}^{n_0} P(n,b)} = \alpha \]

where $P(n,\mu)$ is the Poisson function, $e^{-\mu} \mu^n / n! \;$,
$b$ is the estimated background, 
$\alpha = 0.1$ for a~90\%~CL, and,
since we are computing the ``detector sensitivity'',
$n_0$ is equal to the closest integer number to~$b$. \\

  For each nucleon decay channel we have computed the $(\tau / B)$ limits
as a function of the exposure.
The detection signal efficiency ($\epsilon$) 
and the expected background at each exposure is used to compute the corresponding upper limit ($S$).

\begin{figure}[htbp]
\begin{center}
\vspace{-0.7cm}
\begin{tabular}{c}
 \epsfysize=11.0cm\epsfxsize=11.0cm
 \hspace*{0.1cm}\epsffile{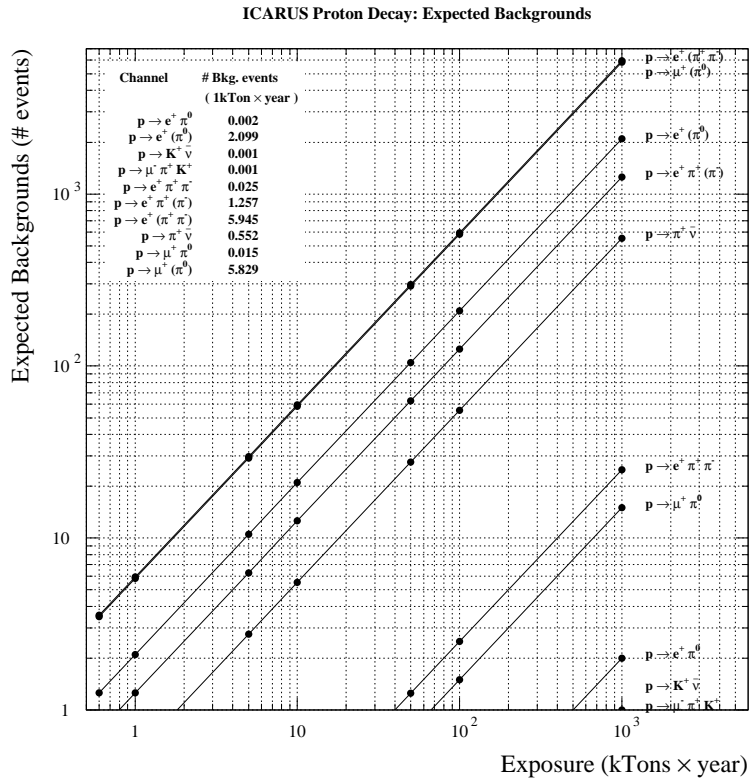} \\
 \epsfysize=11.0cm\epsfxsize=11.0cm
 \hspace*{0.1cm}\epsffile{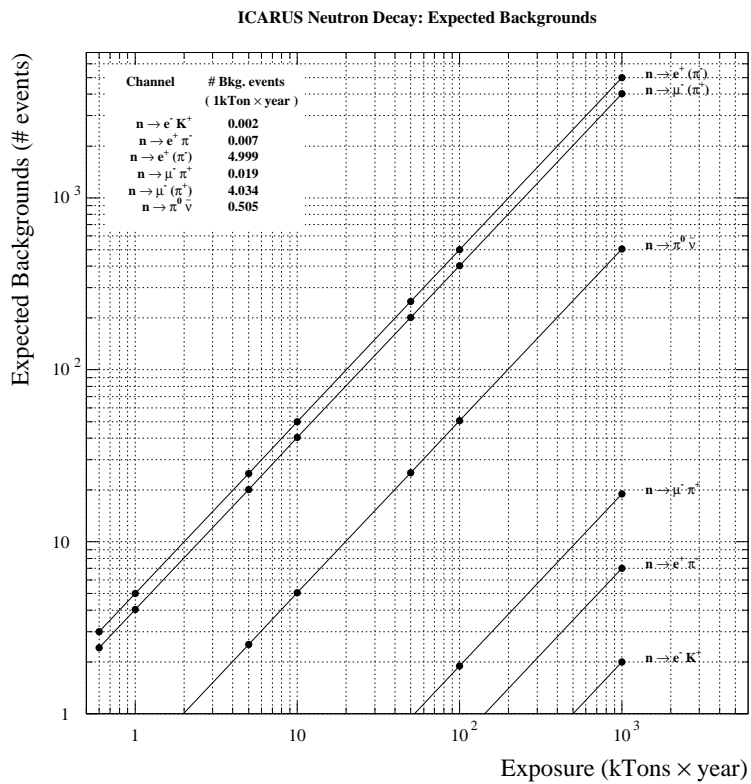}
\end{tabular}
\vspace{0.0cm}
 \caption{\small
Estimated number of background events as a function of the exposure
for the different proton (top) and neutron (bottom) decay channels.
The tables in the plots give the precise values for an
exposure of 1~kton$\times$year.}
 \label{fig:limit_pdk_bkg}
\end{center}
\end{figure}

  The top (bottom) part of figure~\ref{fig:limit_pdk_bkg} shows the
predictions on the number
of background events as a function of the exposure, for the
proton (neutron) decay channels. The tables on the corner of each plot
give the values at 1~kton$\times$year.
As expected, the exclusive channels have much less background than
the inclusive ones. The gap between the two groups (a factor $\sim10^2$)
can be clearly seen in the figure (compare for instance \pepinopi\ and \pepipi ). \\

  From an experimental point of view, this result has two important
consequences. First of all, we observe that there are channels with a
moderate expected contamination already at exposures of
$\sim$1~kton$\times$year. Clearly, the operation of a T600 module, at the
Gran Sasso laboratory, will be of paramount importance to verify the detector efficiencies. 
On the other hand, we find channels that are almost background
free up to exposures of 1~megaton$\times$year. This confirms ICARUS
strength to detect proton decay at the one-event level.

\begin{figure}[htbp]
\begin{center}
\vspace*{-0.7cm}
\begin{tabular}{c}
 \epsfysize=11.0cm\epsfxsize=11.0cm
 \hspace*{0.1cm}\epsffile{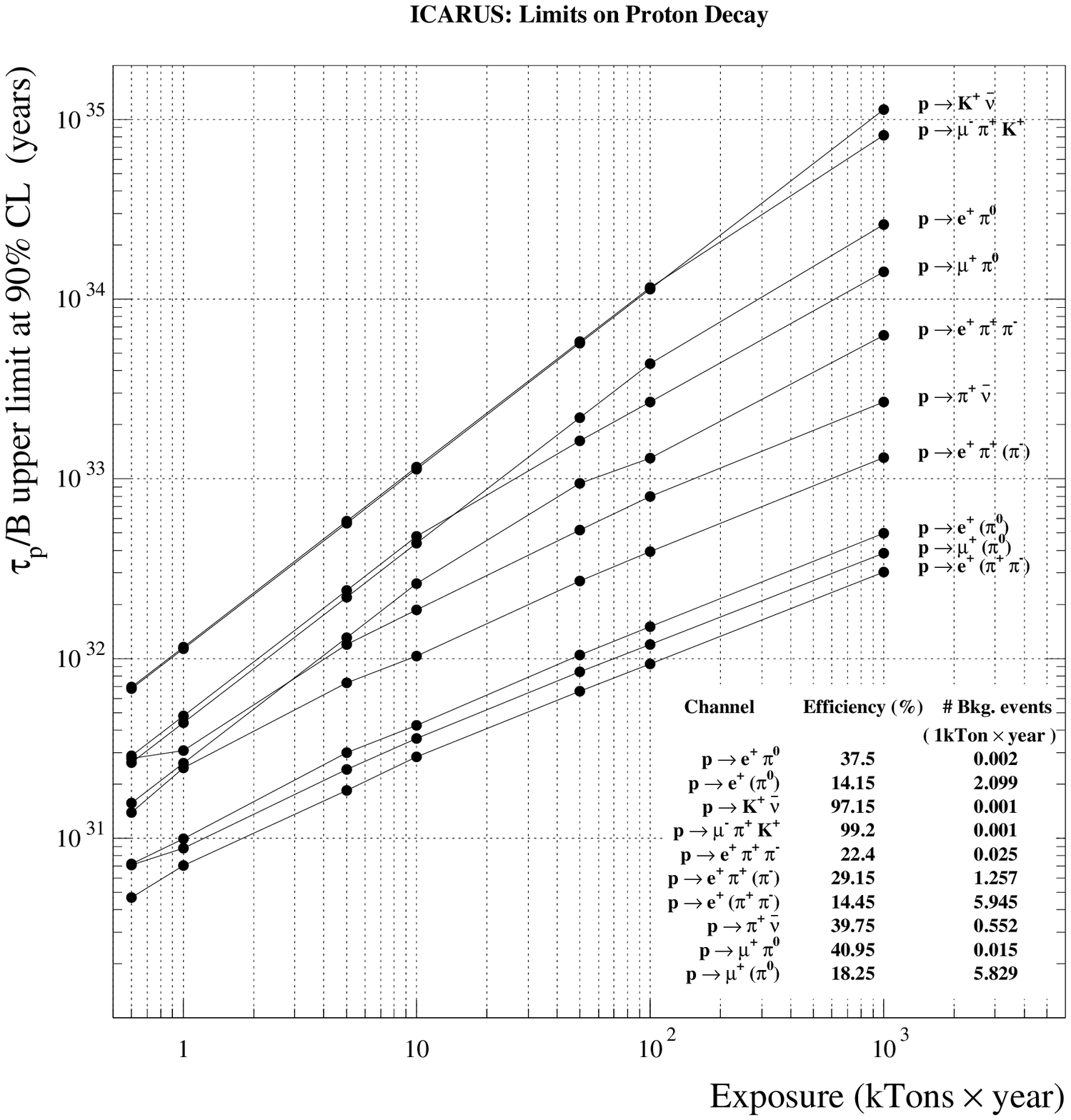} \\
 \epsfysize=11.0cm\epsfxsize=11.0cm
 \hspace*{0.1cm}\epsffile{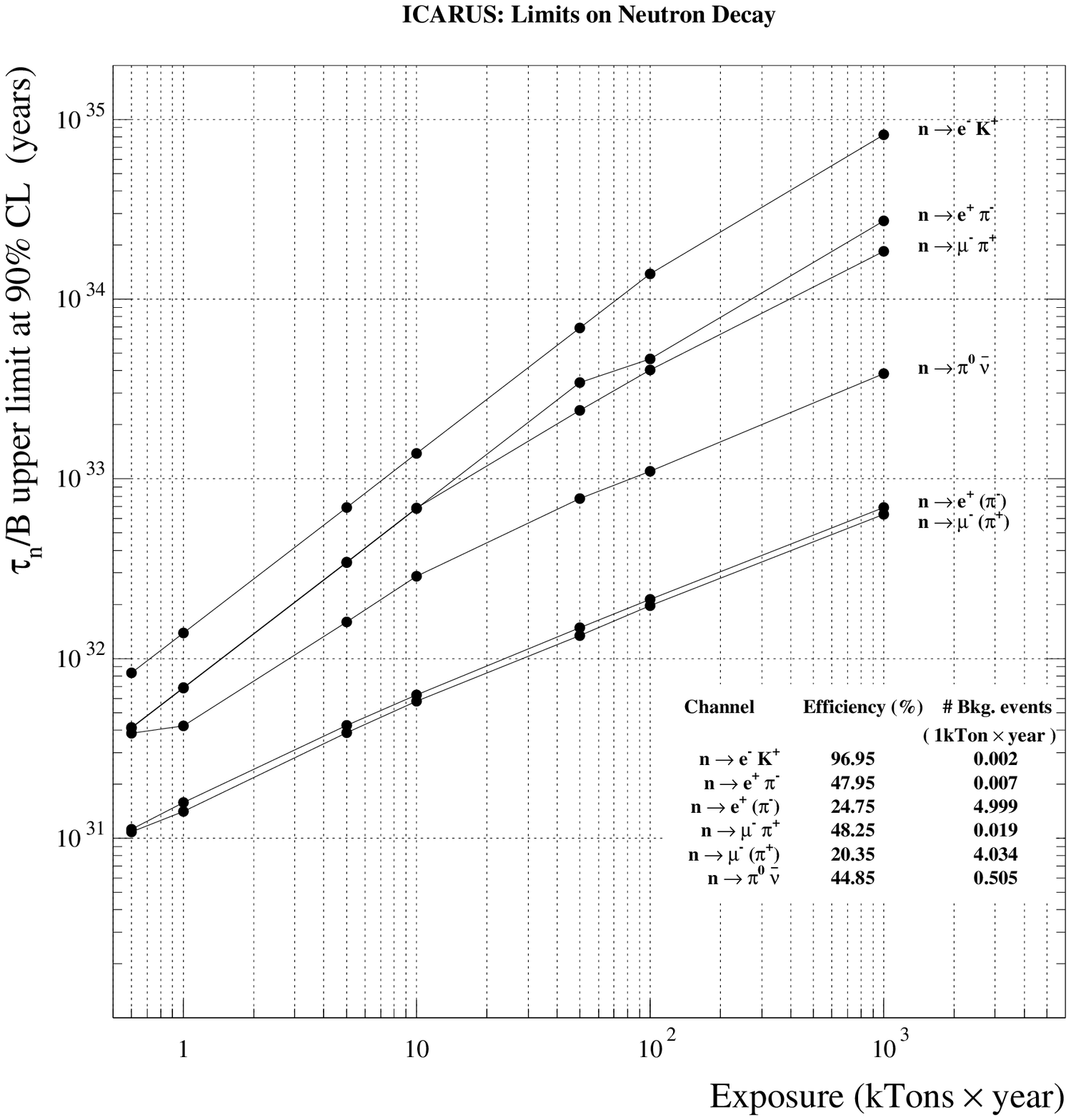}
\end{tabular}
\vspace{0.0cm}
 \caption{\small
Running of the proton (top) and neutron (bottom) decay lifetime limits
($\tau / B$) with the exposure.
The limits are at 90\% confidence level.
The tables indicate the selection efficiencies and the
estimated number of background events for each decay mode, at an exposure
of 1~kton$\times$year.}
 \label{fig:limit_pdk_expo}
\end{center}
\end{figure}

  The obtained proton (neutron) limits on $(\tau / B)$ as a function of
the exposure are illustrated on the top (bottom) part of 
figure~\ref{fig:limit_pdk_expo}.
The tables on the plots indicate the signal efficiencies and the
estimated number of background events for an exposure of
1~kton$\times$year.
In general, the better limits are obtained on the exclusive channels.
The tag based on the presence of a kaon
or a pion accompanied by a charged lepton, with total measured energy
around the proton mass, is powerful enough to annihilate the background. \\

\begin{figure}[htbp]
\begin{center}
\vspace*{-0.7cm}
\begin{tabular}{c}
 \epsfysize=11.0cm\epsfxsize=11.0cm
 \hspace*{0.1cm}\epsffile{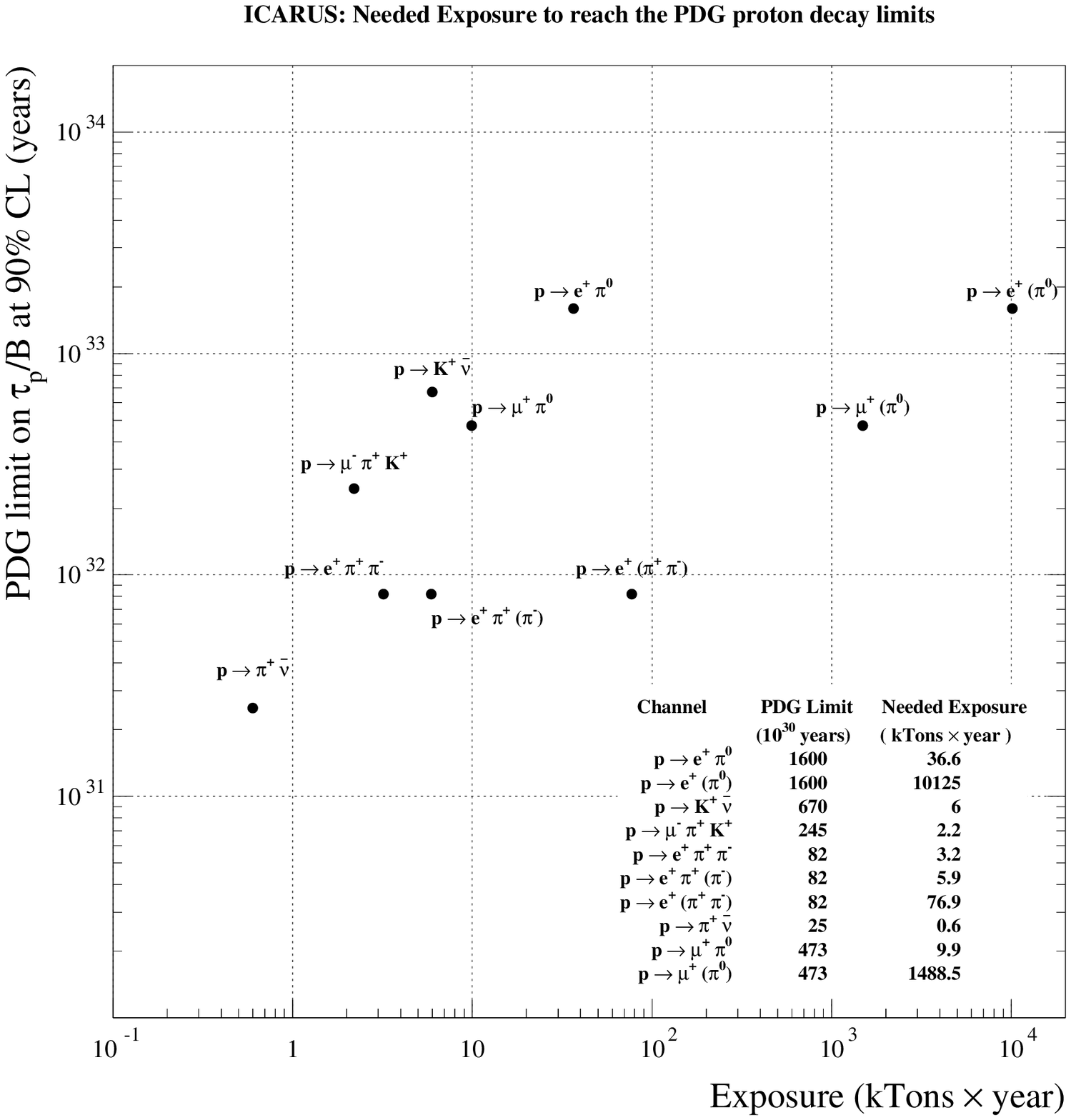} \\
 \epsfysize=11.0cm\epsfxsize=11.0cm
 \hspace*{0.1cm}\epsffile{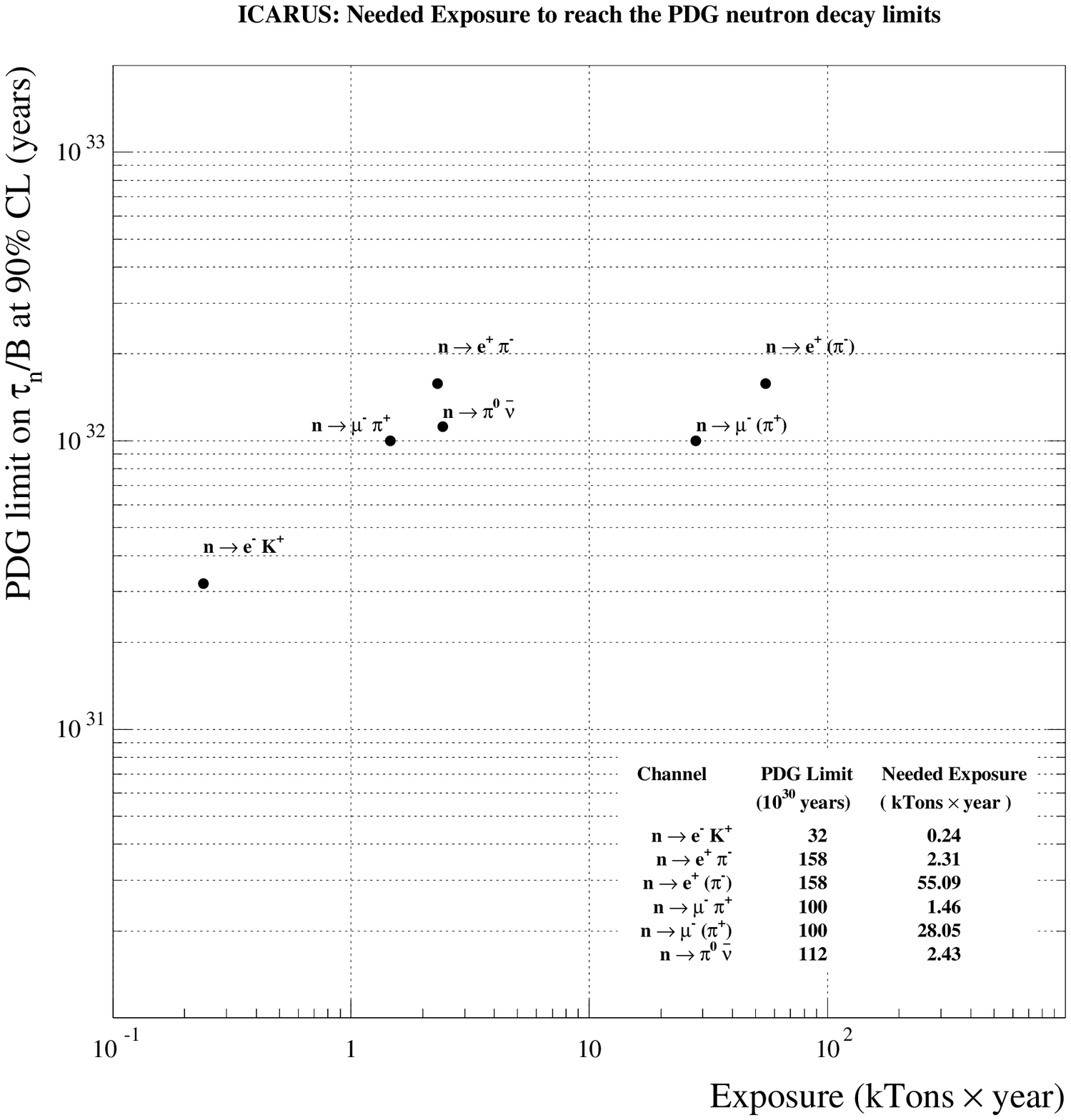}
\end{tabular}
\vspace{0.0cm}
 \caption{\small
Needed exposure (in kilotons$\times$year) to reach the current
PDG proton (top) and neutron (bottom) decay limits.
The tables indicate, for each decay mode,
the precise values of the PDG limits used
and the obtained values of the needed exposure.}
 \label{fig:limit_pdk_pdg}
\end{center}
\end{figure}

  Finally, we have studied how the previous limits on $(\tau / B)$
compare with the current PDG limits~\cite{PDG}.
Figure~\ref{fig:limit_pdk_pdg} shows what would be the minimum exposure
needed to reach the present PDG limits on the different
proton (top) and neutron (bottom) decay modes. 
Both, the used PDG values and the obtained exposures, are
listed on the plots.
It is important to remark that the limits reported in the PDG
refer only to the exclusive channels, so the comparison shown in
figure~\ref{fig:limit_pdk_pdg} is only strictly correct for these channels.
Nevertheless, we also found it interesting to show, how the ICARUS inclusive
limits compare with the PDG exclusive ones.
The modest mass provided by a T600 module is clearly insufficient to
improve existing limits for a vast majority of the reported decay
channels. However, we observe that two years of running will suffice
to increase the sensitivity in channels like, for example, \ppianu\ and
\nek.

The Super-Kamiokande collaboration has recently presented 
preliminary results coming from 70 $kt\times year$
detector exposure~\cite{sk-pdk2000}.
In particular for the p$\rightarrow $e$^+\pi^0$ mode the proton lifetime has been found 
to be higher than $4.4\times$10$^{33}$ years,
while  for p$\rightarrow\bar\nu$K$^+$ the new bound is  1.9$\times$10$^{33}$ 
years at  90$\%$ c.l. The current results show that
while a background-free search can be extended for 
modes like p$\rightarrow $e$^+\pi^0$ and p$\rightarrow \mu^+\pi^0$, 
non-zero background is already expected for the $p\rightarrow \nu K^{+} $ search.
In a water Cherenkov detector the kaon is  below threshold for light emission
and low-background signatures can be obtained only at the 
expense of low efficiency cuts. In particular  
detection of low-energy gamma rays (6.3 MeV) is crucial to background reduction for 
this mode.
This requirement calls for a large photo-cathode coverage in a future detector
based on this technique. Thus only a third 
of the planned half a megaton UNO detector ~\cite{Jung:1999jq}
is going to be dedicated for p$\rightarrow\bar\nu$K$^+$ search.
Consequently a potential for sensitivity improvement for the 
most favored SUSY modes is limited with water Cherenkov technique and 
a liquid argon detector may provide a unique opportunity to either
discover a nucleon decay signal or to rule out 
a large class of theoretical models~\cite{pdk:bpw,pdk:dermisek00}.

\section{Supernova neutrinos}
\label{sec:supernova}

\subsection{Characteristics of a supernova collapse}
For the first few seconds, the gravitational core collapse of a single star 
emits an intense burst of neutrinos whose luminosity rivals the total optical 
emissions of the observable universe. The detection of 20 $\bar\nue$ events
from SN1987A (12 in Kamiokande~\cite{snkamioka} and 8 in IMB~\cite{snimb})
most economically constrained the properties of neutrino
mixing,
neutrino masses, neutrino magnetic moment, neutrino decay, etc.
Those 20 events, however,
could not identify any of the dynamical characteristics of the supernova
mechanism that are stamped up in its neutrino signatures: shock break-out,
convection, accretion, explosion, core cooling, and transparency. Nor
was it possible that any $\numu$ or $\nutau$ be detected, despite the
prediction that these neutrino species and their anti-particles carry
away the bulk of the neutron star binding energy. 

Since then, new experiments have started taking data. For example, the
collection of hundreds to thousands of events is anticipated by the
dedicated LVD experiment. Such information will resolve the many
outstanding questions in supernova modeling, as well as measure or
constrain
the properties of all three generations of neutrinos more tightly than can
now be done in a laboratory. The contribution of ICARUS can provide pieces
of the puzzle not duplicated by other experiments.
 
In order to examine the response of ICARUS to a supernova collapse 
in our Local Group of galaxies, we employ here a generic but detailed model 
of neutrino emission that reflects the latest calculations in Type II supernova 
theory. This baseline model of luminosities and spectra for each of the 
neutrino species incorporates various generic features of the dynamics of 
stellar collapse, but is not tied to any particular model. This patchwork model 
has been presented in Ref.~\cite{burrows92} specifically for the purpose of studying and 
comparing the response and sensitivity of various neutrino detectors. In Figure~\ref{fig:sn1a}
we reproduce from Ref.~\cite{burrows92} the structure of the $\nue$, $\bar\nue$, and ``$\numu$'' 
luminosity curves for the first second of emission, where ``$\numu$'' denotes the 
muon and tau neutrinos and their antiparticles collectively; Figure~\ref{fig:sn1b} shows 
the same for the first 50 seconds. 
The total energy radiated by this model 
is $3 \times 10^{53} erg$, which corresponds to
$2.8\times 10^{57} \ \nu_e$, $1.9\times 10^{57} \ \bar\nu_e$ and 
$4.9\times 10^{57} \ \nu_{\mu,\tau}+\bar\nu_{\mu,\tau}$ for a total 
of  $9.6\times 10^{57}$ neutrinos.
The integrated average 
energies are 11~MeV, 16~MeV, and 25~MeV, respectively.

\subsection{The neutrino signal}
\begin{figure}[htbp]
\centering
\epsfig{file=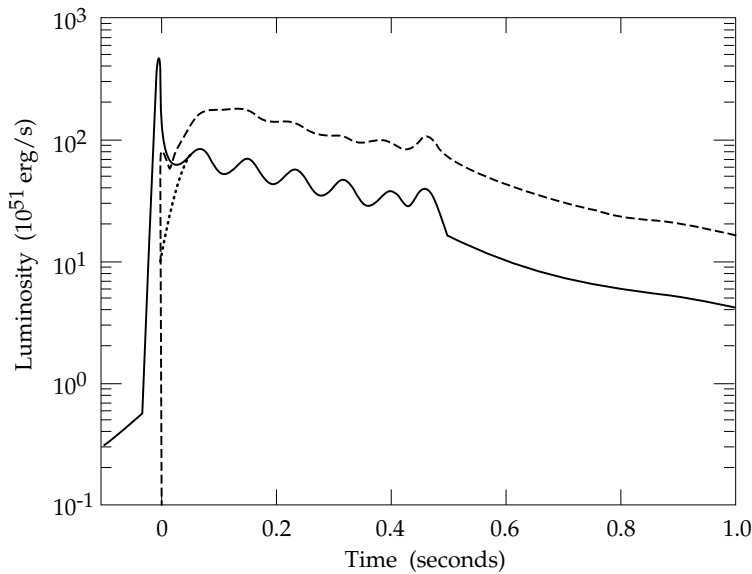,width=10cm}
\caption{Luminosity curves for $\nue$ (solid), $\bar\nue$ (dots), and
``$\numu$'' (dashes) 
for the first second of emission.}
\label{fig:sn1a}
\end{figure}

\begin{figure}[htbp]
\centering
\epsfig{file=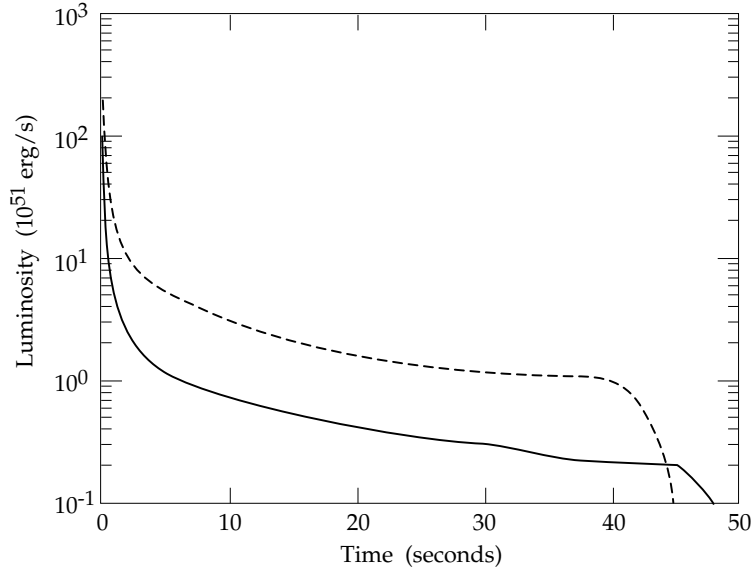,width=10cm}
\caption{Luminosity curves for $\nue$ (solid), $\bar\nue$ (dots), and
``$\numu$'' (dashes) 
for the first 50 seconds of emission.}
\label{fig:sn1b}
\end{figure}

From Figures~\ref{fig:sn1a} and \ref{fig:sn1b}
it is clear that the neutrino emissions are rich in 
diagnostic features of core collapse dynamics and neutron star formation. 
Here we list some of the main features only, for the purpose of their 
identification in the detected neutrino signals. Details may be found in the 
abundant literature; see, for example, Refs.\cite{burrows90,mayle87,woosley86,bethe90}. 
In Figure~\ref{fig:sn1a} the $\nue$ ramp at 
times less than zero is indicative of the accelerating rate of electron capture in 
the collapsing core. The rebound of the inner core into the supersonic outer 
core creates a strong shock wave ($t = 0$), accompanied by the $\nue$ 
neutronization burst (the spike of Figure~\ref{fig:sn1a}), and the sudden turn-on of the 
$\bar\nue$ and ``$\numu$'' radiation. 
 
Immediately after the burst, the $\nue$ spectrum makes a sudden 
transition to a more thermal distribution, with an average 
energy of about 10 MeV. The $\bar\nue$ spectrum is harder, with an average 
energy of 16 MeV. Since the muon and tau neutrinos and their antiparticles 
interact at these temperatures only via neutral current interactions, their 
neutrinosphere lies somewhat deeper within the core. Their spectra is 
therefore hotter than either the $\nue$ or $\bar\nue$ spectra, about 25 MeV. Within 20~ms 
of bounce, hydrostatic equilibrium is achieved and a protoneutron star is 
formed; the $\nue$ and $\bar\nue$ luminosities now merge. At this stage the 
protoneutron star is fattened by accretion of the outer core matter. Either the 
shock wave continues into the outer stellar envelope or it stalls, only to be 
revived within hundreds of milliseconds or seconds by neutrinos from the 
core; these are the so-called prompt and delayed mechanisms, respectively, of 
core collapse. In this patchwork model, the bounce shock fizzles into an 
accretion shock, and subsequent neutrino heating of the shocked envelope 
re-energizes the shock into a supernova at 450~ms. 
 

Oscillations in the mass accretion rate modulate the neutrino 
luminosities, as indicated in Figure~\ref{fig:sn1a}, between 30 and 420 ms. At 450 ms the 
explosion which causes the ejection of the outer envelope and the optical 
supernova display occurs. Accretion shock may delay the explosion by
0.5~seconds 
to some several seconds; its inclusion here serves to display a 
possible structure of neutrino emission, rather than being a necessary 
prediction of the Standard Model. 
Quick spectral hardening, on the other 
hand, is predicted to accompany explosion 
whether it is 
prompt or delayed. While the first 100 ms are rich in diagnostic structure, the 
protoneutron star then begins a long cooling phase which may account for 
most of the energy emitted. During cooling the neutrino luminosities decay 
smoothly according to power laws which reflect the nonlinearity of neutrino 
transport. The long duration is a consequence of the high densities and high 
neutrino energies, which imply high opacities in the protoneutron star 
interior. It is expected that the thousands of events detected from such a 
collapse be spread over many tens of seconds to a minute or more, but that as 
the neutrino energies soften a larger fraction of the emitted luminosity will be 
shunted below detector thresholds. 
Finally, the cooling phase ends when the 
core becomes transparent to neutrinos and the luminosities plummet ($t = 46$ 
and $42$ s in the model used here for $\nue$($\bar\nue$) and ``$\numu$'' respectively). Again, 
since the ``$\numu$'' opacities are lower than the $\nue$ and $\bar\nue$ opacities, it is 
predicted that the ``$\numu$'' emissions should fall off first.
 
\subsection{Event rate in the ICARUS detector}

Models of type II supernovae predict that neutrinos are emitted with
a thermal spectrum, with a temperature hierarchy among neutrino
flavors: $T_{\nu_e} < 
T_{\bar\nu_e} <T_{\nu_\mu,\nu_\tau,\bar\nu_\mu,\bar\nu_\tau}$. 
The neutrino energy spectra can be described by a
Fermi-Dirac distribution~\cite{Langanke:1996he}:
\begin{equation}
\frac{dN}{dE_\nu} = \frac{C}{T^3}\frac{E_\nu^2}{1 + e^{(E_\nu/T-\eta)}} N_\nu
\end{equation}
where $C=0.55$; $T$ is the temperature (MeV); $E_\nu$ 
the neutrino energy (MeV); $\eta$ the chemical potential and
$N_\nu$ the number of expected neutrinos of a given
species. We assume $\eta=0$~\cite{Langanke:1996he}. The
following values for the different neutrino temperatures and 
average energies are used:
\begin{center}
\begin{tabular}{rlcl} 
$\nu_e$ & $T=3.5$ MeV & $\Rightarrow$ & $<E> = 11$ MeV \\
$\bar\nu_e$ & $T=5.0$ MeV & $\Rightarrow$ & $<E> = 16$ MeV \\
$\nu_{\mu,\tau}$ & $T=8.0$ MeV & $\Rightarrow$ & $<E> = 25$ MeV \\
$\bar\nu_{\mu,\tau}$ & $T=8.0$ MeV & $\Rightarrow$ & $<E> = 25$ MeV \\
\end{tabular}
\end{center}

We assume the supernova occurs at 10 kpc, a distance which includes 
53\% of the stars in the galactic disk~\cite{bahcallsn80}. 
All stars in the Milky Way lie within 
30 kpc of the Earth. 
To calculate the theoretical count 
rates for this particular supernova in the ICARUS detector, we convolute the 
luminosities of the various $\nu$ species as a function of time
(Figure~\ref{fig:sn1b}) 
with the 
neutrino scattering and absorption cross-sections on liquid Argon 
and with the average energy spectra as a function of time.

In ICARUS two reactions contribute to the total rate: 
\begin{itemize}
\item {\bf Elastic scattering:} $\nu_x + e^- \to \nu_x + e^- \ (x=e,\mu,\tau)$ 
sensitive to all neutrino species. 
\item {\bf Absorption:} $\nu_e + ^{40}Ar \to e^- + ^{40}K^*$, 
(super-allowed Fermi and Gamow-Teller (GT) transitions~\cite{Ormand:1995js} 
are possible).
\end{itemize}

The elastic neutrino scattering off electrons has a total cross
section that increases linearly with energy:
\begin{eqnarray}
\begin{tabular}{lccc}
$\sigma(\nu_e e^- \to \nu_e e^- )$ & = & $9.20 \times 10^{-45} E_{\nu_e} {\rm
(MeV) \ \ \ \ cm}^2$ \\
$\sigma(\bar\nu_e e^-  \to \bar\nu_e e^- )$ & = & $3.83 \times 10^{-45} E_{\bar\nu_e} {\rm (MeV) \ \ \ \ cm}^2$ \\
$\sigma(\nu_{\mu ,\tau} e^-  \to \nu_{\mu ,\tau} e^- )$ &=& $1.57 \times 10^{-45} E_{\nu_{\mu ,\tau}} {\rm (MeV) \ \ cm}^2$ \\
$\sigma(\bar\nu_{\mu ,\tau} e^-  \to \bar\nu_{\mu ,\tau} e^- )$ &=& $1.29 \times 10^{-45} E_{\bar\nu_{\mu ,\tau}} {\rm (MeV) \ \ cm}^2$ \\
\end{tabular}
\end{eqnarray}

All neutrino species contribute to elastic scattering. The
experimental signature consists of a single recoil electron. Since,
the direction of this electron is highly correlated to the incoming
$\nu$ direction, these events have the potentiality of precisely
determining the location of the supernova source.

\begin{table}[htbp]
\centering
\begin{tabular}{clcccc} \hline
& & & &\multicolumn{2}{c}{Expected events} \\
Reaction & & T (MeV) & $<E_\nu>$ (MeV) & 0.6 ktons & 1.2 ktons \\ \hline
{\bf Elastic} & & & & & \\
& $\nu_e \, e$ & 3.5 & 11 & 4 & 8\\
& $\bar\nu_e\, e$ & 5 & 16 & 2 & 4\\
& $(\nu_\mu+\nu_\tau)\, e$ & 8 & 25 & 1 & 2\\
& $(\bar\nu_\mu +\bar\nu_\tau)\, e$ & 8 & 25 & 1 & 2\\
& total $\nu\, e$ & & & 8 & 16 \\
\hline
{\bf Absorption} & & & & & \\
& $\nu_e\, Ar$ (Fermi) & 3.5 & 11 & 15 & 30\\
& $\nu_e\, Ar$ (GT) & 3.5 & 11 & 30 & 60\\
\hline
{\bf Total} & & & & 53 & 106 \\\hline
\end{tabular}
\caption{Expected neutrino rates for a supernova at a 
distance of 10 kpc, releasing $3\times 10^{53}$ ergs of
binding energy. No energy threshold for electron detection has been applied.}
\label{tab:rates}
\end{table}

Table~\ref{tab:rates} shows the expected rates for this 
reaction. To compute the number of events we fold the total cross
section as a function of energy with the appropriate Fermi-Dirac
distribution. Our rates are calculated integrating over all electron
recoil energies. Since the neutrino burst occurs in a time window of
about 10 seconds, we estimate that the background expected due to
natural radioactivity is negligible and therefore we do not apply 
any detection threshold for electrons. 

\begin{figure}[htbp]
\centering
\epsfig{file=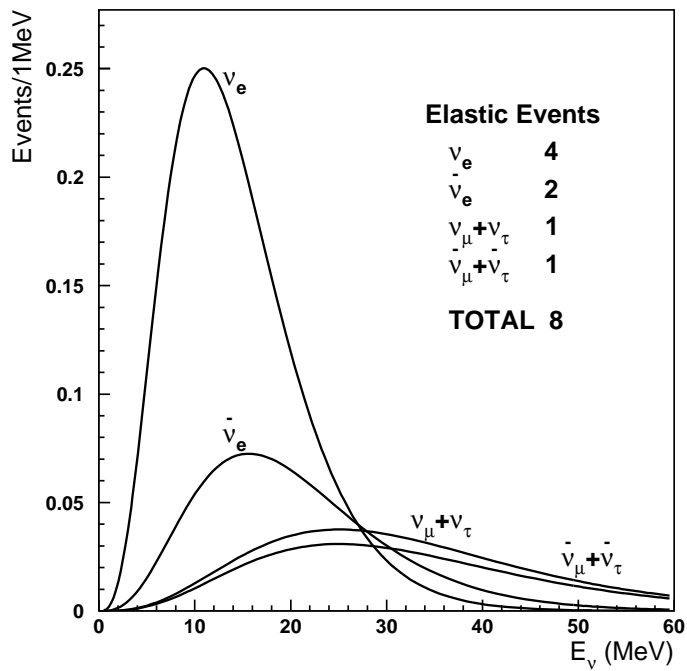,width=10.cm}
\caption{Expected elastic neutrino rates for a type II supernova at a 
distance of 10 kpc. Rates are integrated over all possible electron energies.}
\label{fig:rates}
\end{figure}

Figure~\ref{fig:rates} shows the expected elastic event rates,
as a function of the incoming neutrino
energy, for the case of a 600 tons detector and a supernova occurring at 
10 kpc. The largest contribution ($\sim 50\%$) to elastic events 
comes from $\nu_e$, since they have both the larger flux and cross 
section. For a 0.6 (1.2) kton detector, we expect a total of 8 (16) elastic
events. 

The absorption rate is expected to proceed through two main channels:
a superallowed Fermi transition to the 4.38 MeV excited isobaric
analog K$^*$ state; Gamow-Teller transitions to several excited K
states. The two processes can be distinguished by the energy and
multiplicity of the $\gamma$ rays emitted in the de-excitation and by
the energy spectrum of the primary electron.

The cross section for the Fermi $\nu_e$ capture is given
by~\cite{Raghavan:1986hv}: 
\begin{equation}
\sigma = 1.702 \times 10^{-44} E_e \sqrt{E_e^2-m_e^2} F(E_e) \ \ {\rm cm^2}
\end{equation} 
where the Fermi function, $F(E_e)$, has a value of 1.56 for electron 
energies above 0.5 MeV. The prompt electron energy is 
$E_e = E_\nu + Q - m_e$; $Q$ is the energy threshold ($Q = 5.885$ MeV)
and $m_e$ is the electron mass. We consider
that the GT transition has a cross section which is a factor two 
larger than the one quoted for Fermi absorption~\cite{Ormand:1995js}.

The absorption cross section is larger than for neutrino electron
elastic scattering, hence this process significantly enhances the
sensitivity of ICARUS as a supernova neutrino
detector. Table~\ref{tab:rates} shows the expected rates in case of
$\nu_e$ absorption in Ar. In a 0.6 (1.2) kton detector, we expect 
around 45 (90) absorption events from a supernova located at 10 kpc from Earth.

\begin{figure}[htbp]
\centering
\epsfig{file=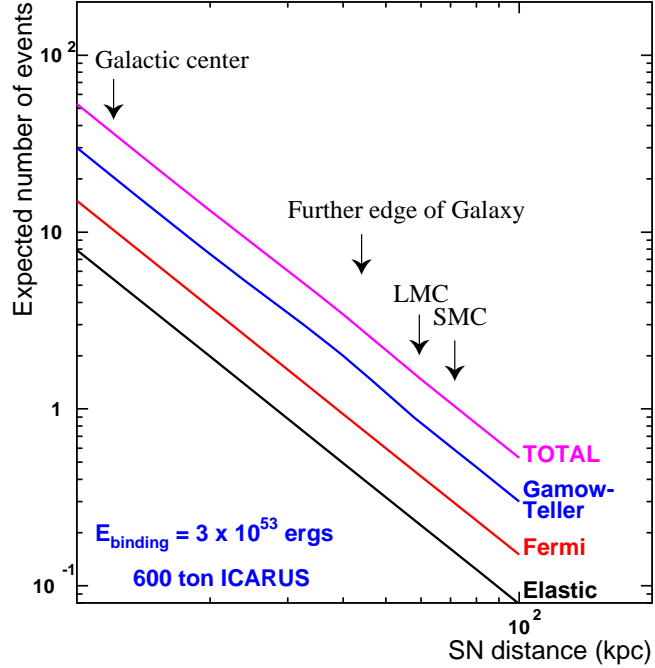,width=10.cm}
\caption{Predicted number of  neutrino events 
for elastic and absorption reactions 
as a function of the supernova distance. We consider a 600 ton detector.}
\label{fig:dist}
\end{figure}

\begin{figure}[htbp]
\centering
\epsfig{file=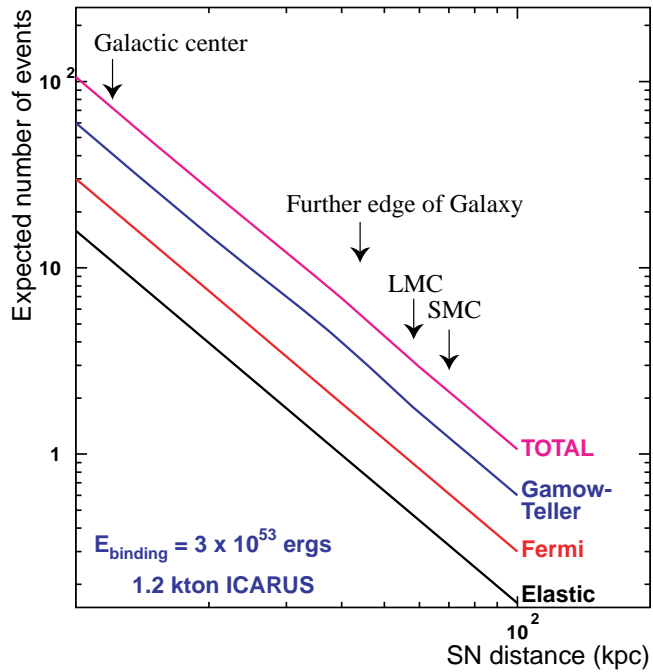,width=10.cm}
\caption{Predicted number of  neutrino events 
for elastic and absorption reactions 
as a function of the supernova distance. We consider a 1.2 kton detector.}
\label{fig:dist12}
\end{figure}

Figures~\ref{fig:dist} and~\ref{fig:dist12} show  the total rate 
for a 0.6 and 1.2 kton detector as a function of supernova distance. It also
displays the expected rates for the  different neutrino reactions. 
For a gravitational stellar collapse occurring in the Large Magellanic
Cloud (distance $\sim$ 60 kpc), we expect to collect, in a 0.6 (1.2) 
kton detector, 2 (4) events as a result of the neutrino burst.

The rich event yield of the first 
second allows many of the characteristics of the supernova mechanism to 
stand out clearly: the $\nue$ neutronization burst, the rapid turn-on of the $\bar\nue$ 
and ``$\numu$'' radiation, the accretion shock oscillations, the explosion pulse, and 
the long cooling of the core. 
The $\nue$ burst is of particular interest and the 
sensitivity of ICARUS to this feature is paralleled only by LVD and SNO; the 
light-water detectors, such as IMB and Kamiokande II, did not have sufficient 
$\nue$ sensitivity to detect it. The count rate peaks at 1200 Hz (but lasts only 
about 20 ms). 

While special triggering 
will be necessary for readout of such a fast event rate, no problems are 
expected with data acquisition as ICARUS has effectively no dead time. 
Accretion shock modulation of the $\nue$ luminosity, depending on its amplitude 
and period of oscillation, may also be discernible by bunching, or pulses of 
about 5 events. The explosion peak should be easily detected, as well as the 
gradual drop in count rate, on account of ICARUS's low energy threshold. 
The overall electron scattering rate in ICARUS, when compared with 
the event rate of $\nue$ absorption, will provide one of the first direct 
confirmations of the existence of the $\numu$ and $\nutau$ components in the supernova 
emission. The light-water neutrino detectors, on the other hand, are largely 
insensitive to these neutrino species. Another advantage of the neutral-
current sensitivity of ICARUS is the fact that the forward-peaked electron 
scattering events will point back to the direction of the supernova, thus 
providing confirmation of their origin.

 

The time structure of the luminosity of the next detected stellar 
collapse, and the absence or presence of structure within it, will yield a 
wealth of information on the explosion mechanism as well as on neutrino 
properties. ICARUS may contribute with hundreds of events to the 
total sample of thousands 
that will be collected by the international effort. ICARUS's neutral-current 
sensitivity will render a part of its contribution complementary to those of 
other detectors. On account of the good energy and angle resolution of the 
detector, its accurate timing, and effectively zero dead time, these events will 
be rich in information on the energy, timing, angle, and flavour content of the 
explosion. Of particular consequence is the fact that ICARUS can be expected 
to be sensitive to the initial $\nue$ neutronization burst, important for the study of 
effects of a finite neutrino mass, as well as the $\numu$ and $\nutau$ components in the 
supernova emission.







\end{document}